\documentclass[preprint,aps,letterpaper]{revtex4}
\usepackage{graphicx,amsmath}
\begin{document}
\newcommand{\etal}{{\it{et al.}}~}
\newcommand{\bqe}{\begin{equation}}
\newcommand{\eqe}{\end{equation}}
\newcommand{\bqs}{\begin{equation*}}
\newcommand{\eqs}{\end{equation*}}
\newcommand{\bqa}{\begin{eqnarray}}
\newcommand{\barint}{-\hspace{-.45cm}\int}
\newcommand{\barinta}{-\hspace{-.37cm}\int}
\newcommand{\eqa}{\end{eqnarray}}
\newcommand{\bqas}{\begin{eqnarray*}}
\newcommand{\eqas}{\end{eqnarray*}}
\newcommand{\nn}{\noindent}
\newcommand{\no}{\nonumber}
\newcommand{\ssi}{\scriptsize}

\title[Optimizing snake locomotion in the plane. I. Computations]
{Optimizing snake locomotion in the plane. I. Computations}

%\author{Silas Alben$^{1,2}$ and
%Fangxu Jing$^2$}
\author{Silas Alben}

\affiliation{Department of Mathematics, University of Michigan,
Ann Arbor, MI, USA}
\label{firstpage}

\begin{abstract}
We develop a numerical scheme to determine which
planar snake motions are optimal for locomotory efficiency,
across a wide range of frictional parameter space. For a large
coefficient of transverse friction, we find that retrograde
traveling waves are optimal. The optimal snake deflection
scales as the -1/4 power of the coefficient of
transverse friction, in agreement with an asymptotic
analysis. At the other
extreme, zero coefficient of transverse friction, we
propose a triangular direct wave which is optimal. Between
these two extremes, a variety of complex, locally optimal
motions are found. Some of these can be classified as
standing waves (or ratcheting motions).
\\
\\
\\
Keywords: snake; friction; sliding; locomotion; optimization.
\end{abstract}

\maketitle
%\corraddr{(alben@umich.edu)}
\section{Introduction}

Snake locomotion has a long history of study by biologists,
engineers, and applied mathematicians
\citep{gray1950kinetics,GuMa2008a,HuNiScSh2009a,HaCh2010b,MaHu2012a,HuSh2012a}. A lack of appendages
makes snake motions simpler in some respects than those of
other locomoting animals \citep{dickinson2000animals}. However, snakes
can move through a wide range of terrestrial
\citep{gray1946mechanism,gray1950kinetics,jayne1986kinematics},
aquatic \citep{shine2003aquatic}, and aerial
\citep{socha2002kinematics}
environments, with different kinematics corresponding
to the different dynamical laws which apply in each setting.
Here we focus on terrestrial locomotion, with
motions restricted to two dimensions,
and a Coulomb frictional force acting on the snake.
Even with these simplifying assumptions, there are many possible
snake kinematics to consider.
One way to organize our understanding of a diversity of locomotory
kinematics is to propose a measure of locomotory performance (here,
efficiency), and determine which motions are optimal, and
how their performance varies with physical parameters.
Well-known examples are optimization studies of organisms
moving in low- \citep{BeKoSt2003a,AvGaKe2004a,TaHo2007a,fu2007theory,spagnolie2010optimal,crowdy2011two}
and high-Reynolds-number fluid flows
\citep{lighthill1975mathematica,childress1981mechanics,sparenberg1994hydrodynamic,alben2009passive,michelin2009resonance,peng2012bb}.

In this work, we use a recently-proposed
model for snake locomotion in the plane which has shown good agreement
with biological snakes \citep{HuNiScSh2009a,HuSh2012a}. The snake is
a slender body whose curvature changes as a prescribed function
of arc-length distance along the backbone, and time. Its net rotation and translation are then determined by
Newton's laws, with Coulomb friction acting between the snake and the ground. This model is perhaps one of the simplest which can
represent arbitrary planar snake motions. Previous studies have
used the model to find optimal snake motions when the
motions are restricted to sinusoidal traveling waves
\citep{HuSh2012a},
or the bodies are composed of two or three links \citep{JiAl2013}.
These motions are represented
by a small number of parameters (2-10); at the upper end of this
range, the optimization problem is difficult to solve using
current methods. In \citep{GuMa2008a}, Guo and Mahadevan developed a model which included
the internal elasticity and viscosity of the snake, and
studied the effects of these and other physical parameters on
locomotion for a prescribed sinusoidal shape and sinusoidal and
square-wave internal bending moments. In the present work,
we neglect the internal mechanics of the snake, and consider
only the work done by the snake on its environment, the ground.
However, we consider a more general class of snake motions.
Hu and Shelley assumed a sinusoidal traveling-wave
snake shape, and computed
the snake speed and locomotory efficiency across the two-parameter
space of traveling-wave amplitude and wavelength for two values
of frictional coefficients \citep{HuSh2012a}. They also considered the ability of
the snake to redistribute its weight during locomotion.
Jing and Alben considered the locomotion of two- and three-link
snakes, and found analytical and numerical results for
the scaling of snake speed and efficiency \citep{JiAl2013}. Among the results
were the optimal temporal function for actuating the
internal angles between the links, expressed as Fourier
series with one and two frequencies.

Here we address the optimization problem for
planar snakes more generally, by using a much larger number of parameters
(45 in most cases) to represent the snake's curvature as a function
of arc length and time. Although
our snake shape space is limited, it is large enough to represent
a wide range of shapes. To solve the optimization problem
in a 45-dimensional shape space, and simultaneously
across a large region of the two-dimensional friction parameter space,
we develop an efficient
numerical method based on a quasi-Newton method. The simulations
show clearly that two types of traveling-wave motions, retrograde
and direct
waves, are optimal when the coefficient
of friction transverse to the snake is large and small, respectively.
An asymptotic analysis in \cite{AlbenSnake2013II} shows how
the snake's traveling wave amplitude should scale with
the transverse coefficient of friction when it is large. The analysis
gives a $-1/4$ power law for the amplitude which agrees well
with the numerical results shown here. When the transverse
coefficient of friction is zero, we give another
optimally-efficient motion here which is
a direct wave of deflection along the body.
Between these extremes, the numerics show
a more complicated set of optima which
include standing wave (or ratcheting) motions, among a wide
variety of other locally-optimal motions. Taken together,
these results show which planar snake motions are optimal
within a large space of possible motions and frictional
coefficients.

\section{Model \label{sec:Model}}

\begin{figure} [h]
           \begin{center}
           \begin{tabular}{c}
               \includegraphics[width=5in]{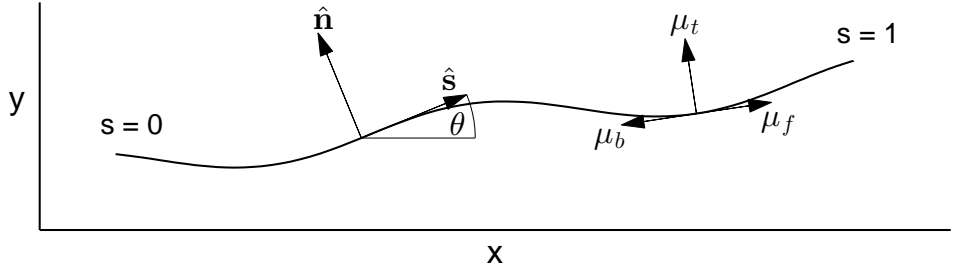} \\
           \vspace{-.25in}
           \end{tabular}
          \caption{\footnotesize Schematic diagram of snake position,
parametrized by arc length $s$ (nondimensionalized by snake length),
 at an instant in time.  The tangent angle
and unit vectors
tangent and normal to the curve at a point are labeled. Vectors representing
forward, backward and transverse velocities are shown with
the corresponding friction coefficients $\mu_f$, $\mu_b$, and $\mu_t$.
 \label{fig:SnakeSchematic}}
           \end{center}
         \vspace{-.10in}
        \end{figure}

We use the same frictional
snake model as \citep{HuNiScSh2009a,HuSh2012a,JiAl2013}, so we only summarize it
here. The snake's
position is given by $\mathbf{X}(s,t) = (x(s,t), y(s,t))$, a planar curve which is
parametrized by arc length $s$ and varies with time $t$. A schematic
diagram is shown in figure \ref{fig:SnakeSchematic}.

The unit vectors
tangent and normal to the curve are $\hat{\mathbf{s}}$ and $\hat{\mathbf{n}}$
respectively. The tangent angle and curvature are denoted $\theta(s,t)$ and $\kappa(s,t)$,
and satisfy $\partial_s x = \cos\theta$, $\partial_s y = \sin\theta$, and
$\kappa = \partial_s \theta$. We consider the problem of prescribing the curvature of
the snake as a function of time, $\kappa(s,t)$, in order to obtain efficient locomotion.
When $\kappa(s,t)$ is prescribed, the tangent angle and position are obtained by integration:
\begin{align}
\theta(s,t) &= \theta_0(t) + \int_0^s \kappa(s',t) ds', \label{theta0} \\
x(s,t) & = x_0(t) + \int_0^s \cos \theta(s',t) ds', \label{x0}\\
y(s,t) &= y_0(t) + \int_0^s \sin \theta(s',t) ds'. \label{y0}
\end{align}
\nn The trailing-edge position $(x_0, y_0)$ and tangent angle $\theta_0$
are determined by the force and torque
balance for the snake, i.e. Newton's second law:
\begin{align}
\int_0^L \rho \partial_{tt} x ds &= \int_0^L f_x ds, \label{fx0} \\
\int_0^L \rho \partial_{tt} y ds &= \int_0^L f_y ds, \label{fy0} \\
\int_0^L  \rho \mathbf{X}^\perp \cdot \partial_{tt} \mathbf{X} ds
&= \int_0^L \mathbf{X}^\perp \cdot \mathbf{f} ds. \label{torque0}
\end{align}
\nn Here $\rho$ is the snake's mass per unit length and $L$ is
the snake length. The snake is locally inextensible,
and $\rho$ and $L$ are constant in time.
$\mathbf{f}$ is the force per unit length on the snake
due to Coulomb friction with the ground \cite{HuNiScSh2009a}:
\bqe
\mathbf{f}(s,t) = -\rho g \mu_t
\left( \widehat{\partial_t{\mathbf{X}}}\cdot \hat{\mathbf{n}} \right)\hat{\mathbf{n}}
-\rho g \left(\mu_f H(\widehat{\partial_t{\mathbf{X}}}\cdot \hat{\mathbf{s}})
+ \mu_b (1-H(\widehat{\partial_t{\mathbf{X}}}\cdot \hat{\mathbf{s}}))\right)
\left( \widehat{\partial_t{\mathbf{X}}}\cdot \hat{\mathbf{s}} \right)\hat{\mathbf{s}}. \label{friction}
\eqe
\nn Here $H$ is the Heaviside function and the hats denote
normalized vectors. When $\|\partial_t{\mathbf{X}}\| = \mathbf{0}$
we define $\widehat{\partial_t{\mathbf{X}}}$ to be $\mathbf{0}$.
According to (\ref{friction}) the snake experiences
friction with different coefficients for motions in different directions.
The frictional coefficients are $\mu_f$, $\mu_b$, and $\mu_t$ for motions
in the forward ($\hat{\mathbf{s}}$), backward ($-\hat{\mathbf{s}}$),
and transverse (i.e. normal) directions ($\pm\hat{\mathbf{n}}$), respectively. In
general the snake velocity at a given point has both tangential and
normal components, and the frictional force density has
components acting in each direction. A similar decomposition of force
into directional components
occurs for viscous fluid forces on slender bodies \citep{cox1970motion}.

We assume that the snake curvature $\kappa(s,t)$ is a prescribed function of
$s$ and $t$ that is periodic in $t$ with period $T$. Many of the motions
commonly observed in real snakes are essentially periodic in time \citep{HuNiScSh2009a}.
We nondimensionalize equations (\ref{fx0})--(\ref{torque0}) by dividing
lengths by the snake length $L$, time by $T$, and mass by $\rho L$. Dividing
both sides by $\mu_f g$ we obtain:
\begin{align}
\frac{L}{\mu_f gT^2} \int_0^1 \partial_{tt} x ds &= \int_0^1 f_x ds, \label{fxa} \\
\frac{L}{\mu_f gT^2}\int_0^1 \partial_{tt} y ds &= \int_0^1 f_y ds, \label{fya} \\
\frac{L}{\mu_f gT^2}\int_0^1 \mathbf{X}^\perp \cdot \partial_{tt} \mathbf{X} ds
&= \int_0^1 \mathbf{X}^\perp \cdot \mathbf{f} ds. \label{torquea}
\end{align}
\nn In (\ref{fxa})--(\ref{torquea}) and from now on, all variables are
dimensionless. For most of the snake motions observed in nature, $L/\mu_f gT^2 \ll 1$
\citep{HuNiScSh2009a}, which means that the snake's inertia is negligible. By setting
this parameter to zero we simplify the problem
considerably while maintaining
 a good representation of real snakes. (\ref{fxa})--(\ref{torquea}) become:
\begin{align}
\mathbf{b} = (b_1,b_2,b_3)^\top = \mathbf{0} \quad ; \quad b_1 &\equiv \int_0^1 f_x ds, \label{fxb} \\
b_2 &\equiv \int_0^1 f_y ds, \label{fyb} \\
b_3 &\equiv \int_0^1 \mathbf{X}^\perp \cdot \mathbf{f} ds. \label{torqueb}
\end{align}
\nn In (\ref{fxb})--(\ref{torqueb}), the dimensionless force $\mathbf{f}$ is
\bqe
\mathbf{f}(s,t) = -\frac{\mu_t}{\mu_f}
\left( \widehat{\partial_t{\mathbf{X}}}\cdot \hat{\mathbf{n}} \right)\hat{\mathbf{n}}
- \left( H(\widehat{\partial_t{\mathbf{X}}}\cdot \hat{\mathbf{s}})
+ \frac{\mu_b}{\mu_f} (1-H(\widehat{\partial_t{\mathbf{X}}}\cdot \hat{\mathbf{s}}))\right)
\left( \widehat{\partial_t{\mathbf{X}}}\cdot \hat{\mathbf{s}} \right)\hat{\mathbf{s}} \label{friction1}
\eqe
\nn The equations (\ref{fxb})--(\ref{torqueb}) thus involve only two parameters,
which are ratios of the friction coefficients. From now on, for simplicity, we
refer to $\mu_t/\mu_f$ as $\mu_t$ and $\mu_b/\mu_f$ as $\mu_b$. Without
loss of generality, we assume $\mu_b \geq 1$. This amounts to defining
the backward direction as that with the higher of the tangential frictional
coefficients, when they are unequal. $\mu_t$ may assume any nonnegative value.
The same model was used in \citep{HuNiScSh2009a,HuSh2012a,JiAl2013},
and was found to agree well with the motions of biological
snakes in \citep{HuNiScSh2009a}.

Given the curvature $\kappa(s,t)$, we solve the three nonlinear equations
(\ref{fxb})--(\ref{torqueb}) at each time $t$ for the three unknowns
$x_0(t)$, $y_0(t)$ and $\theta_0(t)$. Then we obtain the snake's
position as a function of time by (\ref{theta0})--(\ref{y0}).
The distance
traveled by the snake's center of mass over one period is
\bqe
d = \sqrt{\left(\int_0^1 x(s,1) - x(s,0) \,ds\right)^2 +
\left(\int_0^1 y(s,1) - y(s,0)\, ds\right)^2}. \label{dist}
\eqe
The work done by the snake against friction over one period is
\begin{align}
W &= \int_0^1 \int_0^1 \mathbf{f}(s,t) \cdot \partial_t\mathbf{X}(s,t)\, ds\, dt. \label{W}
\end{align}
\nn
% We introduce notation for the tangential and normal components of
% the snake velocity:
% \bqe
% u_s = \partial_t{\mathbf{X}}\cdot \hat{\mathbf{s}} \quad ;
% \quad u_n = \partial_t{\mathbf{X}}\cdot \hat{\mathbf{n}}.
% \eqe
% \nn In this notation
% \bqe
% W = \int_0^1 \int_0^1 \frac{\mu_t u_n^2 +
% \left(H(u_s) + \mu_b(1-H(u_s))\right)u_s^2}{\sqrt{u_s^2 + u_n^2}}\, ds\, dt. \label{W0}
% \eqe
We define the cost of locomotion as
\bqe
\eta = \frac{W}{d} \label{eta}
\eqe
\nn and our objective is to find $\kappa(s,t)$ which minimize $\eta$
at $(\mu_b, \mu_t)$ values which range widely over
$[1,\infty) \times [0,\infty)$. In our computations,
instead of $\eta$, we minimize a different function which is more convenient for
physical and numerical reasons:
\bqe
F = -\frac{d}{W} e^{2\cos(\theta(0,1)-\theta(0,0))}. \label{F}
\eqe
\nn To obtain $F$ from $\eta$ we have substituted
$-d/W$ for $W/d$, to avoid infinities in the objective function
for a common class of motions with $d \to 0$ and $W$ of order 1.
The exponential term in (\ref{F}) penalizes rotations over
a period, to ensure that the snake moves in a straight path
rather than a circular path over many periods. In the Supplementary
Material section \ref{sec:F}
we give a more detailed explanation for the choice of $F$.

% \bqe
% \int \left[\left( \widehat{\partial_t{\mathbf{X}}}\cdot \hat{\mathbf{s}} \right) s_x +
% \mu_t \left( \widehat{\partial_t{\mathbf{X}}}\cdot \hat{\mathbf{n}} \right) n_x \right]\, ds = 0.
% \label{fxc}
% \eqe

\section{Numerical Minimization Method \label{sec:NumMeth}}

Our goal is to determine the snake shape, $\kappa(s,t)$, which
minimizes $F$ for a given $(\mu_b, \mu_t)$. Since
$\kappa(s,t)$ has $t$-period one, we represent it via a
double-series expansion:
\bqe
\kappa(s,t) = \sum_{j = 0}^{m_1-1} \sum_{k = 0}^{n_1-1}
\left(\alpha_{jk} \cos(2\pi j t) + \beta_{jk} \sin(2\pi j t)\right) T_k(s). \label{expn}
\eqe
\nn Here $T_k$ is a Chebyshev polynomial in $s$:
\bqe
T_k(s) = \cos(k \arccos(s)).
\eqe
\nn The expansion (\ref{expn}) converges as $m_1, n_1 \to \infty$
for a large class of $\kappa$ which includes differentiable functions, and
with a convergence rate that increases rapidly
with the number of bounded derivatives of $\kappa$
\citep{boyd2001chebyshev}. It is reasonable to hope
that minimizing $F$ over a class of
functions (\ref{expn}) with small $m_1$ and
$n_1$ gives a good approximation to the minimizers with infinite $m_1$ and
$n_1$. Since the terms in (\ref{expn}) with coefficients $\beta_{0k}$ are automatically
zero, the functions given by (\ref{expn}) are a $(2m_1 -1)n_1$-dimensional space. In
most of this work we use $m_1 = n_1 = 5$, so we are minimizing $F$ in a 45-dimensional
space. We give a few results at large $\mu_t$ with $m_1 = n_1 = 10$ for comparison, and find
similar minimizers in this region of parameter space. The minimization algorithm
converges in a smaller portion of $(\mu_b, \mu_t)$ space as
$m_1$ and $n_1$ increase, and convergence is considerably slowed by
the increased dimensionality of the search space.
We find that $m_1 = n_1 = 5$ is small
enough to allow convergence to minima in a large portion of $(\mu_b, \mu_t)$ space,
but large enough to approximate the gross features of a wide range of
$\kappa(s,t)$. Another reason to expect that good approximations
to minimizers can be obtained with small $n_1$ is that
the dynamical equations (\ref{fxb})--(\ref{torqueb}) depend only
on spatial integrals of body position and velocity.

Using the dynamical equations (\ref{fxb})--(\ref{torqueb}), we
show in Supplementary
Material section \ref{sec:t} that $d$, $W$, and $F$ are the same for any reparametrization of time. Such a reparametrization can be used to
reduce the high-frequency components of a motion while keeping the efficiency
the same. Therefore, it is reasonable to expect that a good approximation to
any minimizer of $F$ can be obtained with low
temporal frequencies, that is, with an expansion in the form of (\ref{expn}) with
small $m_1$.

We use the Broyden-Fletcher-Goldfarb-Shanno (BFGS) algorithm \citep{nocedal1999numerical}
to minimize $F$ numerically, by computing a sequence of iterates which converge
to a local minimizer of $F$. The algorithm requires routines
for computing the values and gradients of $F$ as well as a line search method for
finding the next iterate at each step, using the search direction provided within
the BFGS algorithm. The BFGS algorithm is a quasi-Newton algorithm, and therefore
forms an approximation to the Hessian matrix of second derivatives, without the
expense and difficulty of computing the Hessian matrix explicitly.

We have developed an efficient method for evaluating $F$ and its
gradient with respect to the $(2m_1 -1)n_1$ shape parameters with
just a single simulation of the snake trajectory over one period.
Computing the value of $F$ requires computing the trajectory of the snake over one
period. We discretize the period interval uniformly with $m$ time points:
 $\{0, 1/m, 2/m, \ldots, 1-1/m\}$. At each time point we write
(\ref{fxb})--(\ref{torqueb}) as a nonlinear algebraic system of three equations
in three unknowns: $\{\partial_t x_0, \partial_t y_0, \partial_t \theta_0\}$.
To formulate the system, we set $\{x_0, y_0, \theta_0\}$ to $\{0,0,0\}$ at each
time point, which means that we set the snake in an artificial frame
rotated from the lab frame. $\partial_t \theta_0$ is the same in the two
frames, so after solving for $\{\partial_t x_0, \partial_t y_0, \partial_t \theta_0\}$
in the artificial frame, we integrate $\partial_t \theta_0$ in time
to obtain $\theta_0(t)$ in the lab frame. This function
is the angle between the artificial frame and the lab frame at each
time, and we use it to rotate $\{\partial_t x_0, \partial_t y_0\}$ to the
lab frame. We then integrate $\{\partial_t x_0, \partial_t y_0\}$ in
the lab frame to
obtain $\{x_0(t), y_0(t)\}$, and integrate $\kappa(s,t)$ using
$\{x_0(t), y_0(t), \theta_0(t)\}$ to obtain the body trajectory
{\it a posteriori}. Using $\{\partial_t x_0, \partial_t y_0, \partial_t \theta_0\}$
as the unknowns instead of $\{x_0, y_0, \theta_0\}$ has two
advantages: we avoid cancellation error associated with
computing discrete time derivatives of $\{x_0, y_0, \theta_0\}$,
and we solve $m$ decoupled systems of three equations in three
unknowns instead of a large coupled system of $3m$ equations
in $3m$ unknowns ($\{x_0, y_0, \theta_0\}$ at all times points),
which is more expensive to solve.

We solve the system (\ref{fxb})--(\ref{torqueb}) using Newton's method
with a finite-difference
Jacobian matrix. We solve it at each time sequentially from $t = 0$,
using a random initial guess for
$\{\partial_t x_0, \partial_t y_0, \partial_t \theta_0\}$ at $t=0$
and an extrapolation from previous time points as an initial guess at other $t$.
An important element of our method is the quadrature used to compute the integrals
in (\ref{fxb})--(\ref{torqueb}) using (\ref{friction1}).
Here we introduce notation for the tangential and normal components of
the snake velocity:
\bqe
u_s = \partial_t{\mathbf{X}}\cdot \hat{\mathbf{s}} \quad ;
\quad u_n = \partial_t{\mathbf{X}}\cdot \hat{\mathbf{n}}.
\eqe
In (\ref{friction1}) we can write
\bqe
\widehat{\partial_t{\mathbf{X}}}\cdot \hat{\mathbf{s}} =
\frac{u_s}{\sqrt{u_s^2 + u_n^2}} \quad ; \quad
\widehat{\partial_t{\mathbf{X}}}\cdot \hat{\mathbf{n}} =
\frac{u_n}{\sqrt{u_s^2 + u_n^2}}.
\eqe
\nn These terms have unbounded derivatives with respect to
$u_s$ and $u_n$ when $u_s, u_n \to 0$. The local error
when integrating (\ref{friction1}) using the trapezoidal rule (for example)
with uniform mesh size $h$ is $O\left(h^3\,\mbox{max}(u_s,u_n)^{-2}\right)$.
To obtain convergence in this case it is necessary to have
$h = o(\mbox{max}(u_s,u_n)^{2/3})$, so
the trapezoidal rule (and other classical quadrature rules)
needs to be locally adaptive when $(u_s, u_n)$ becomes small.

We use a different approach which allows for a uniform mesh even
when $(u_s, u_n) \to 0$.
% The idea is that the
% If the integrals are computed
% using the trapezoidal rule with $\partial_t\mathbf{X}$
% discretized on a uniform mesh with spacing $h$, $O(h)$ absolute errors are made
% on intervals where $\|\partial_t \mathbf{X}(s,t)\| = 0$. Much larger (unbounded)
% errors are made in the finite-difference Jacobian matrix, because the quadrature
% formula has unbounded derivatives with respect to
% $\{\partial_t x_0, \partial_t y_0, \partial_t \theta_0\}$. With this
% quadrature
% Newton's method does not converge for typical shape functions $\kappa(s,t)$.
We perform the quadrature analytically on subintervals using locally
linear approximations for $u_s$ and $u_n$.
% In (\ref{friction1}) we write
% \bqe
% \widehat{\partial_t{\mathbf{X}}}\cdot \hat{\mathbf{s}} =
% \frac{u_s}{\sqrt{u_s^2 + u_n^2}} \quad ; \quad
% \widehat{\partial_t{\mathbf{X}}}\cdot \hat{\mathbf{n}} =
% \frac{u_n}{\sqrt{u_s^2 + u_n^2}}.
% \eqe
% \nn
Using $\{\partial_t x_0, \partial_t y_0, \partial_t \theta_0\}$ and
$\kappa(s,t)$, we compute $\partial_t \mathbf{X}$, $\mathbf{X}$,
and then $u_s$ and $u_n$ on a uniform mesh. On each subinterval,
we approximate $u_s$ and $u_n$ as linear functions
of the form $A s+B$ using
their values at the interval endpoints. We
approximate $\mathbf{X}^\perp$, $\hat{\mathbf{s}}$, and $\hat{\mathbf{n}}$ as constants using
their values at the midpoints of the intervals. Then the
integrals in (\ref{fxb})--(\ref{torqueb}), using (\ref{friction1}),
are of the form
\bqe
C_0 \int_a^b \frac{A s + B}{\sqrt{(As + B)^2 + (Cs + D)^2}}\,ds \label{form}
\eqe
\nn on each subinterval. We evaluate such integrals analytically using
\begin{align}
I_1 &\equiv \int_0^1 \frac{ds}{\sqrt{s^2 + a_2 s + a_3}} = -\log\left(a_2 + 2\sqrt{a3}\right)
+ \log\left(2 + a_2 + 2\sqrt{1+a_2 + a_3}\right),\label{I1} \\
I_2 &\equiv \int_0^1 \frac{s\,ds}{\sqrt{s^2 + a_2 s + a_3}} =
-\sqrt{a_3} + \sqrt{1 + a_2 + a_3} -\frac{a_2}{2}I_1. \label{I2}
\end{align}
\nn When used to evaluate (\ref{form}),
the logarithms in (\ref{I1})--(\ref{I2}) have canceling
singularities when the linear approximations to
$u_s$ and $u_n$ are proportional (including the case in which one is zero),
i.e. when $AD-BC \to 0$. When this occurs, we use different formulae
because the integrands are constants or sign functions,
which are easy to integrate analytically.
This quadrature method is $O(h^2)$ for mesh size $h$,
due to the use of linear and midpoint approximations of the functions
in the integrands.

With this method for evaluating (\ref{fxb})--(\ref{torqueb}), we compute
$\{\partial_t x_0, \partial_t y_0, \partial_t \theta_0\}$ by Newton's method
as described above,
and obtain the snake trajectory and $d$, $W$, and $F$ for
the prescribed curvature function.

The BFGS method also requires the gradient of $F$ with
respect to the $(2m_1 -1)n_1$ curvature coefficients
$\{\alpha_{pq}, \beta_{pq}\}$. For notation, let
$\mathbf{a}$ be a $(2m_1 -1)n_1 \times 1$ vector
whose entries are the curvature coefficients.
We compute a finite-difference gradient using
\bqe
\nabla F_j = \frac{F(\mathbf{a} +
10^{-8}\|\mathbf{a}\|\mathbf{e}_j) - F(\mathbf{a}))}
{10^{-8}\|\mathbf{a}\|}, \; j = 1, \ldots, (2m_1 -1)n_1, \label{grad}
\eqe
\nn where $\mathbf{e}_j$ is the unit basis vector
in the $j$th coordinate direction.
(\ref{grad}) requires $(2m_1 -1)n_1 + 1$
evaluations of $F$, or $(2m_1 -1)n_1 + 1$
computations of the snake motion over a period. We
employ a very accurate approximation which requires
only a single computation of the snake motion over a period.

Computing $F(\mathbf{a} + 10^{-8}\|\mathbf{a}\|\mathbf{e}_j)$
requires solving $\mathbf{b} = \mathbf{0}$, i.e. (\ref{fxb})--(\ref{torqueb}),
for $\{\partial_t x_0, \partial_t y_0, \partial_t \theta_0\}$
at each time $t_k = k/m \in [0, 1/m, \ldots 1-1/m]$
using the vector of curvature coefficients
$\mathbf{a} + 10^{-8}\|\mathbf{a}\|\mathbf{e}_j$.
$\mathbf{b}$ is a function of
$(\partial_t x_0, \partial_t y_0, \partial_t \theta_0)^\top$
and the curvature coefficients, so we write
it as $\mathbf{b}((\partial_t x_0, \partial_t y_0, \partial_t \theta_0)^\top,
\mathbf{a} + 10^{-8}\|\mathbf{a}\|\mathbf{e}_j)$ and solve
$\mathbf{b}((\partial_t x_0, \partial_t y_0, \partial_t \theta_0)^\top,
\mathbf{a} + 10^{-8}\|\mathbf{a}\|\mathbf{e}_j) = \mathbf{0}$
for $(\partial_t x_0, \partial_t y_0, \partial_t \theta_0)^\top$
at each time $t_k$. The value of
$(\partial_t x_0, \partial_t y_0, \partial_t \theta_0)^\top$
which satisifies these equations at each $t_k$
may be written
$\mathbf{c}_k(\mathbf{a} + 10^{-8}\|\mathbf{a}\|\mathbf{e}_j)$.
Having simulated the snake motion and evaluated $F$ when
the curvature coefficient vector is $\mathbf{a}$, we
have $\mathbf{c}_k(\mathbf{a})$ for each $t_k$.
$\mathbf{c}_k(\mathbf{a})$ solves $\mathbf{b}((\partial_t x_0, \partial_t y_0, \partial_t \theta_0)^\top,
\mathbf{a}) = \mathbf{0}$, a set of equations whose terms are
extremely close to (within about $10^{-8}$ of) those of
$\mathbf{b}((\partial_t x_0, \partial_t y_0, \partial_t \theta_0)^\top,
\mathbf{a} + 10^{-8}\|\mathbf{a}\|\mathbf{e}_j) = \mathbf{0}$
at each $t_k$. Thus $\mathbf{c}_k(\mathbf{a})$ is an
extremely good initial guess for
$\mathbf{c}_k(\mathbf{a} + 10^{-8}\|\mathbf{a}\|\mathbf{e}_j)$,
and we need only perform a single Newton iteration with
this initial guess to obtain
$\mathbf{c}_k(\mathbf{a} + 10^{-8}\|\mathbf{a}\|\mathbf{e}_j)$
to machine accuracy (about $10^{-16}$).
The one-step Newton iteration is as follows.
We define $\mathbf{\epsilon_j} \equiv 10^{-8}\|\mathbf{a}\|\mathbf{e}_j$ for
our description. We start with the usual Taylor series expansion from
which Newton's method is obtained:
\bqe
0 =
\mathbf{b}(\mathbf{c}_k(\mathbf{a} + \mathbf{\epsilon_j}), \mathbf{a} + \mathbf{\epsilon_j})
= \mathbf{b}(\mathbf{c}_k(\mathbf{a}), \mathbf{a} + \mathbf{\epsilon_j}) +
\underline{\underline{J}}(\mathbf{c}_k(\mathbf{a}), \mathbf{a} + \mathbf{\epsilon_j})
\left(\mathbf{c}_k(\mathbf{a} + \mathbf{\epsilon_j}) - \mathbf{c}_k(\mathbf{a})\right)
+ O(\|\mathbf{\epsilon_j}\|^2). \label{Newt1}
\eqe
\nn Here $\underline{\underline{J}}$ is the 3-by-3 Jacobian matrix of partial
derivatives of $\mathbf{b}$ with respect to its leading 3-by-1 vector argument:
\bqe
\underline{\underline{J}}(\mathbf{x},\mathbf{y})_{ij} \equiv
\partial_{x_j} b_i(\mathbf{x},\mathbf{y}), \quad i = 1, 2, 3, \; j = 1, 2, 3.
\eqe
\nn Having previously simulated the snake motion with the curvature
coefficient vector $\mathbf{a}$ to evaluate $F(\mathbf{a})$,
we have computed $\mathbf{c}_k(\mathbf{a})$ via a
separate instance of Newton's method, and
we have also calculated and stored
$\underline{\underline{J}}(\mathbf{c}_k(\mathbf{a}), \mathbf{a})$ at
each $t_k$. We therefore modify $\underline{\underline{J}}$
in (\ref{Newt1}) to use this information, while retaining
the same order of accuracy in $\mathbf{\epsilon_j}$:
\bqe
0 =
\mathbf{b}(\mathbf{c}_k(\mathbf{a} + \mathbf{\epsilon_j}),
\mathbf{a} + \mathbf{\epsilon_j})
= \mathbf{b}(\mathbf{c}_k(\mathbf{a}), \mathbf{a} + \mathbf{\epsilon_j}) +
\underline{\underline{J}}(\mathbf{c}_k(\mathbf{a}), \mathbf{a})
\left(\mathbf{c}_k(\mathbf{a} + \mathbf{\epsilon_j}) - \mathbf{c}_k(\mathbf{a})\right)
+ O(\|\mathbf{\epsilon_j}\|^2). \label{Newt2}
\eqe
\nn We solve (\ref{Newt2}) to obtain $\mathbf{c}_{jk}$, our approximation to
$\mathbf{c}_k(\mathbf{a} + \mathbf{\epsilon_j})$ using a single Newton iteration:
\bqe
\mathbf{c}_{jk} \equiv  \mathbf{c}_k(\mathbf{a}) -
\underline{\underline{J}}(\mathbf{c}_k(\mathbf{a}), \mathbf{a})^{-1}
\mathbf{b}(\mathbf{c}_k(\mathbf{a}), \mathbf{a} + \mathbf{\epsilon_j})
= \mathbf{c}_k(\mathbf{a} + \mathbf{\epsilon_j}) + O(\|\mathbf{\epsilon_j}\|^2).
\label{ctilde}
\eqe
\nn (\ref{ctilde}) requires only a single
computation of the snake motion over a period, which
gives $\mathbf{c}_k(\mathbf{a})$ and
$\underline{\underline{J}}(\mathbf{c}_k((\mathbf{a}), \mathbf{a})^{-1}$
at each $t_k$. It also requires $(2m_1 -1)n_1$ evaluations of
the function $\mathbf{b}$. These are much less expensive than even a
single computation of the snake motion over a period. The
$O(\|\mathbf{\epsilon_j}\|^2)$ error in $\mathbf{c}_{jk}$ is of the same
order of magnitude as machine precision. The
value of the cost function
$\tilde{F}_j$ computed from
$\mathbf{c}_{jk}$ is approximately within
machine precision of $F(\mathbf{a} + \mathbf{\epsilon_j})$
(which corresponds to $\mathbf{c}_k(\mathbf{a} + \mathbf{\epsilon_j})$), and therefore
$\tilde{F}_j$ contributes an error of the same order as the $\approx 10^{-8}$ finite-difference
error in (\ref{grad}) when it is used in place of
$F(\mathbf{a} + \mathbf{\epsilon_j})$ in (\ref{grad}).
This level of error leads to essentially no deterioriation
in the convergence of BFGS up to machine precision, relative
to the version with the exact gradient
\citep{kelley1987iterative}. Therefore we obtain the
gradient of $F$ for approximately the cost of computing $F$.
Each computation of a search direction in
BFGS requires only one gradient computation. These ingredients give
an efficient optimization method, which is critical to finding
optima in the high-dimensional space of curvature coefficients.

\section{Large-$\mu_t$ results}

\begin{figure} [h]
           \begin{center}
           \begin{tabular}{c}
               \includegraphics[width=5in]{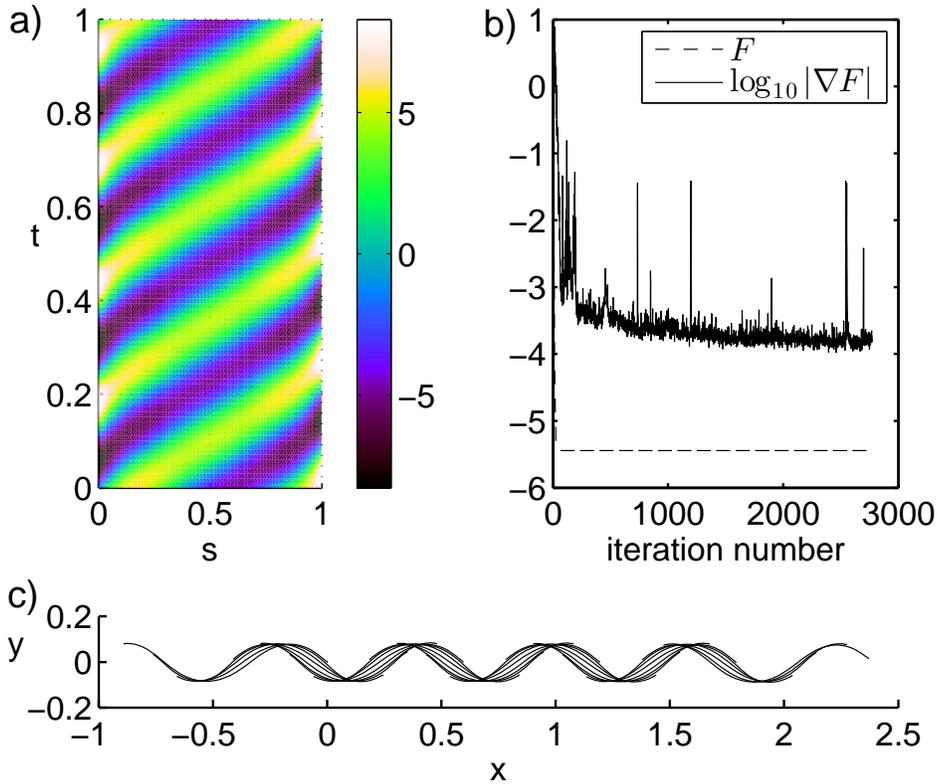} \\
           \vspace{-.25in}
           \end{tabular}
          \caption{\footnotesize Result of one snake optimization run for $\mu_b = 1$ and $\mu_t = 30$.
(a) Map of curvature versus arc length and time. (b) Values of the objective function
(dashed line) and $\log_{10}$
of the 2-norm of its gradient (solid line)
with respect to the the iteration count in the optimization routine.
(c) Snake trajectory over one period.
 \label{fig:FirstSinusoidalFig}}
           \end{center}
         \vspace{-.10in}
        \end{figure}

In Figure \ref{fig:FirstSinusoidalFig} we show the result of one run of the optimization
routine with $\mu_b = 1$ and $\mu_t = 30$, starting with a randomly-chosen
initial vector of curvature coefficients. In panel a we plot a map of curvature values
$\kappa(s,t)$ over one period, and
find that the curvature approximates a sinusoidal traveling wave
with wavelength approximately
equal to the snake body length. In this case the waves appear to travel
with nearly constant speed, shown by the straightness of the curvature bands. The bands deviate
from straightness near the ends, where the largest curvatures occur. In panel
b we plot the
objective function and $\log_{10}$ of the 2-norm of its gradient over the iteration count
of the optimization routine. We find that the gradient norm decreases by four orders of magnitude
within the first 100 iterations, and the objective function reaches a plateau at the same time.
In panel c we give snapshots of the position of the snake $\mathbf{X}(s,t)$ at equal instants in
time, moving from left to right, and see that the snake approximately
follows a sinusoidal path.

\begin{figure} [h]
           \begin{center}
           \begin{tabular}{c}
               \includegraphics[width=6in]{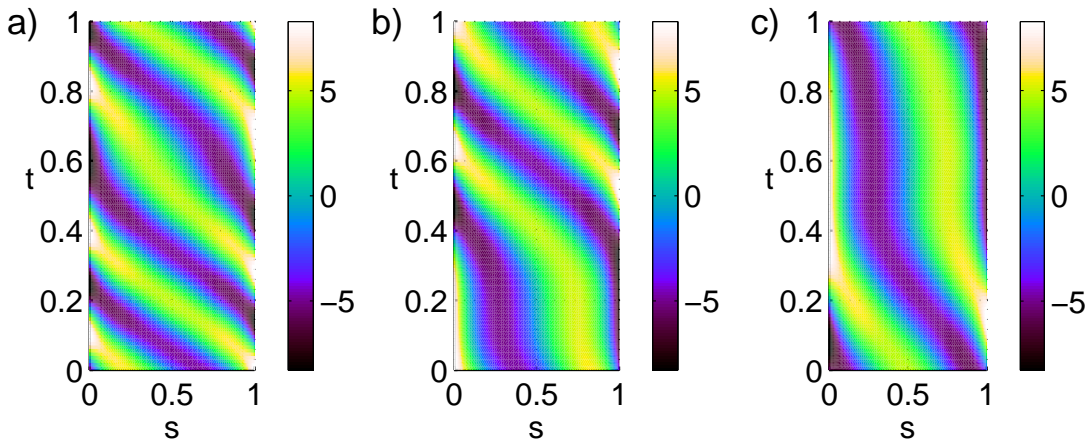} \\
           \vspace{-.25in}
           \end{tabular}
          \caption{\footnotesize Results of three additional snake
optimization routines for $\mu_b = 1$ and $\mu_t = 30$.
(a-c) Maps of curvature versus arc length and time after convergence, starting from three different
randomly-chosen initial iterates.
 \label{fig:AltSinusoidalFig}}
           \end{center}
         \vspace{-.10in}
        \end{figure}

Figure \ref{fig:AltSinusoidalFig} shows three different converged results for the same
parameters $\mu_b = 1$ and $\mu_t = 30$ but with different
randomly-chosen initial iterates. Again we have traveling waves of curvature
shown by bands, but unlike in Figure \ref{fig:FirstSinusoidalFig}a, the bands are not
straight. Also, the bands travel from right to left instead of left to right. The bands are
not straight because the traveling wave speed changes over time. However, at any given
time, the distances in $s$ from the maximum to the minimum of curvature
are nearly the same in
Figure \ref{fig:AltSinusoidalFig}a--c and Figure \ref{fig:FirstSinusoidalFig}a, and
so are the shapes.
In all four cases, the snake passes through the same sequence of shapes, but
at different speeds, and different (integer) numbers of times in one period.
Thus the four motions are essentially the same after
a reparametrization of time. Correspondingly, the values of $F$ for all four are in the range
$-5.4418 \pm 0.0024$. The distances traveled, $d$, are within 0.2\% of integer multiples
of 0.597, with the multiples 4,3,2, and 1 for the motions in Figure \ref{fig:FirstSinusoidalFig}a and
Figures \ref{fig:AltSinusoidalFig}a--c, respectively, which correspond to the number of
times they move through the sequence of shapes represented by Figure \ref{fig:AltSinusoidalFig}c.
The values of $F$ are the same whether the waves move from left to right or right
to left in $s$, because $\mu_b$ (or $\mu_b/\mu_f) = 1$, so there is a symmetry between
forward and backward motions.

Over an ensemble of 20 random initializations, this sinusoidal motion yields the lowest
value of $F$ among the local optima found.
We have repeated the search with twice as many spatial and temporal modes ($m_1$ = 10,
$n_1$ = 10) and 48 random initializations. About three-quarters of the
searches converge, and
the majority of the converged states are
very close to those shown in Figures \ref{fig:FirstSinusoidalFig}
and \ref{fig:AltSinusoidalFig}. The values of $F$ for these states lie in the range
$-5.577 \pm 0.026$. The mean is systematically lower than the mean for $m_1$ = 5,
$n_1$ = 5 by about 2.5\%. The decrease in the objective function is slight, and
the optimal shape is fairly smooth, which indicates that the $m_1$ = 5,
$n_1$ = 5 minimizer is representative of the minimizer for much larger $m_1$ and $n_1$,
at least for $\mu_b = 1$ and $\mu_t = 30$. With $m_1$ = 10 and $n_1$ = 10,
the curvature bands of
Figures \ref{fig:FirstSinusoidalFig}a and \ref{fig:AltSinusoidalFig} become
straighter near the ends, showing the effect of the truncation in the Chebyshev series.
However, it is more difficult to obtain convergence with
four times as many modes, because of the higher dimensionality of the space. Since a
primary goal here is to obtain the optima across as broad a range of ($\mu_b,\mu_t$)
space as possible, we proceed with $m_1$ = 5 and $n_1$ = 5.

\begin{figure} [h]
           \begin{center}
           \begin{tabular}{c}
               \includegraphics[width=5in]{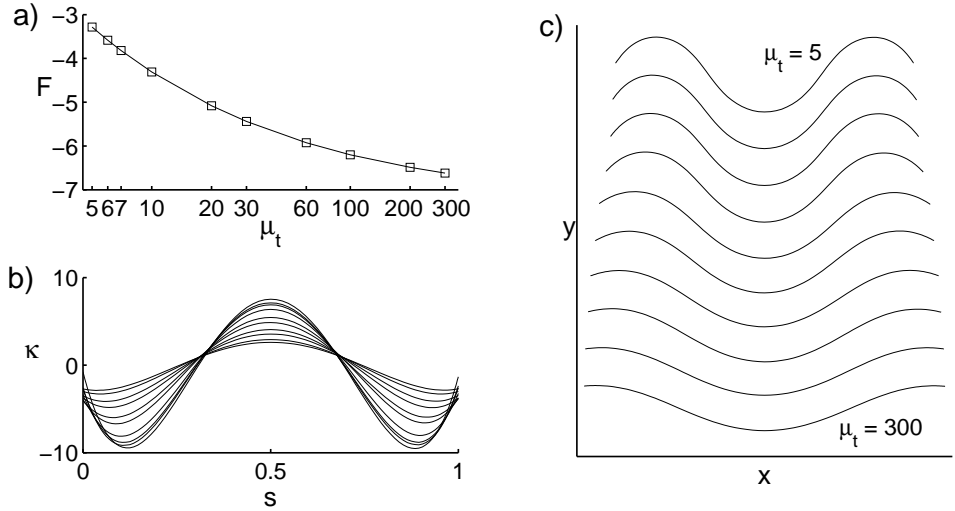} \\
           \vspace{-.25in}
           \end{tabular}
          \caption{\footnotesize Traveling-wave optima
for ten different $\mu_t$ on a logarithmic scale.
a) Values of $F$. b) Curvature versus
arc length at the instant when a curvature maximum crosses the
midpoint, for the ten $\mu_t$ values in (a). c) Snake shapes corresponding to (b).
 \label{fig:HighMutFig}}
           \end{center}
         \vspace{-.10in}
        \end{figure}

Performing searches with $\mu_b = 1$ and a wide range of $\mu_t$, we
find that similar sinusoidal
traveling-wave optima occur for $\mu_t \gtrsim 6$.
In Figure \ref{fig:HighMutFig}a we plot the values of $F$ corresponding
to the optima over a range of $\mu_t$ on a logarithmic scale,
and find that the optima are more efficient at higher $\mu_t$. In
panel b, the curvatures of the optimal shapes are plotted for
the same $\mu_t$, at the times when the curvatures have a local maximum
at $s = 1/2$. The corresponding snake body positions are shown in
panel c, and are vertically displaced to make different bodies
easier to distinguish. In all cases the bodies approximate
sinusoidal shapes with slightly more than one wavelength
of curvature on the body. The curvature amplitudes decrease
monotonically with increasing $\mu_t$.

Performing searches with a wide range of $\mu_t$ \textit{and} $\mu_b$ now,
we find that similar sinusoidal traveling-wave optima are found for
$\mu_t \gtrsim 6$ and all $\mu_b$ used
($1 \leq \mu_b \leq 3000$). For these motions,
the curvature traveling waves move backward in $s$,
and the tangential motion is purely forward,
so they have the same $F$ at all $\mu_b$. However,
for $\mu_b > 1$, a separate class of (local, not global) optima are also found.
These have
curvature traveling waves which move \textit{forward} in $s$,
and for which the tangential motion is purely backward. Since friction is
higher in the backward direction, these local optima have higher $F$
(are less efficient) than the oppositely-moving optima, and
the values of $F$ for these optima increase with increasing $\mu_b$,
for the same reason.

\begin{figure} [h]
           \begin{center}
           \begin{tabular}{c}
               \includegraphics[width=6in]{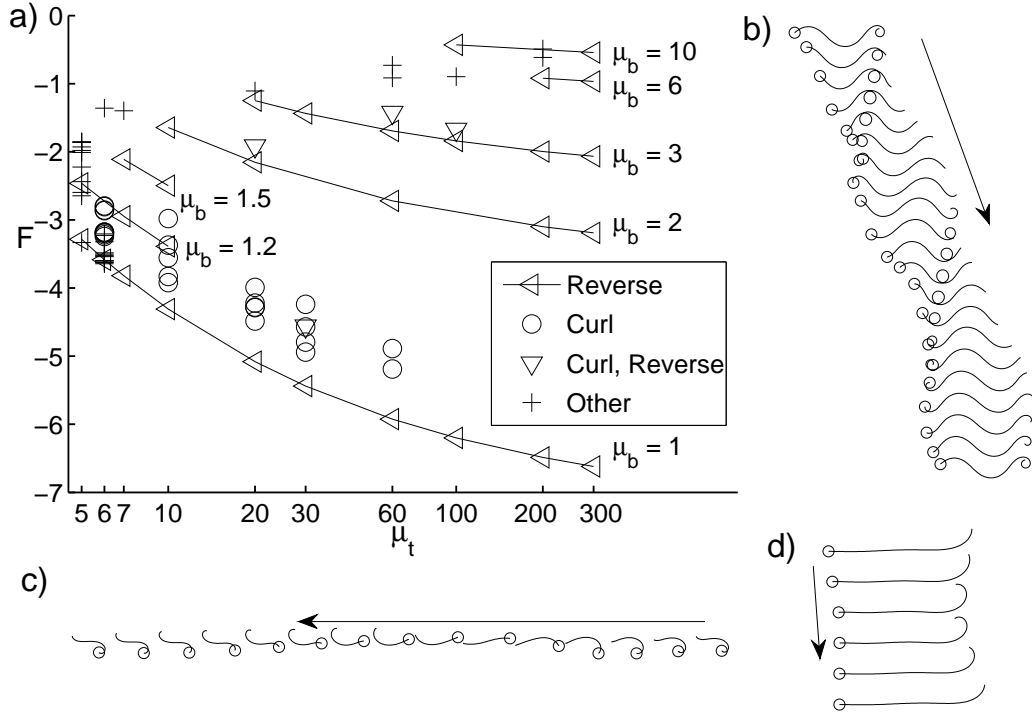} \\
           \vspace{-.25in}
           \end{tabular}
          \caption{\footnotesize Four categories of locally optimal motions:
reverse sinusoidal motion,
sinusoidal motion with curls, reverse sinusoidal motion with curls, and other motions,
for ten different $\mu_t$ on a logarithmic scale.
a) Values of $F$. Corresponding values of $\mu_b$ are labeled only for the
reverse sinusoidal motions. b) Snapshots of a reverse sinusoidal motion with curls. Here
$\mu_b = 1$ and $\mu_t = 10$.
c,d) Snapshots of motions in the ``Other'' category, with $\mu_b = 10$ and $\mu_t = 60$
for (c) and $\mu_b = 6$ and $\mu_t = 20$ for (d).
 \label{fig:HighMutReverseFig}}
           \end{center}
         \vspace{-.10in}
        \end{figure}

In figure \ref{fig:HighMutReverseFig}a we plot the values of $F$ for these
``Reverse'' sinusoidal motions at various $\mu_b$ and $\mu_t$, with
leftward-pointing triangles. The triangles connected by a given
solid line are data for various $\mu_t$ at
a particular value of $\mu_b$, labeled
to the right of the line. When $\mu_b = 1$, the reverse sinusoidal
motions have the same $F$ values as the forward sinusoidal motions shown
in figure \ref{fig:HighMutFig}. These have the lowest $F$ values
across all $\mu_b$ for $\mu_t \gtrsim 6$, and
are thus the most efficient motions we have found.

Three other sets of symbols are used to denote other classes of
local optima found by our method. All are less efficient than the
forward sinusoidal motions, but we present examples of them to show
some of the types of local optima that exist for this problem. For these
three sets, shown by circles, downward pointing triangles, and plus
signs, we do not join the optima with a given $\mu_b$ by lines, to reduce
visual clutter. The circles, labeled ``Curl,''
give $F$ values for shapes which are like sinusoidal traveling waves,
but they have a curl---the snake body winds around itself once
in a closed loop. Snapshots of the snake motion for such
a case are shown in figure \ref{fig:HighMutReverseFig}b.
Here the snake moves horizontally, but the snapshots are displaced vertically
to make them distinguishable. The arrow shows the direction of motion.
The tail of the snake is marked by a circle. The snake body moves
rightward while the curvature wave and curl both move leftward.
Because the snake overlaps itself, this motion is not physically
valid in 2D. However, a large variety of such optima are found numerically,
with various small-integer numbers of curls
(and with both directions of rotation). The efficiency of such
motions decreases as the number of curls increases.
The downward-pointing triangles, labeled ``Curl, Reverse,'' are for
traveling waves with curls which move in the forward direction along the snake.
These have a yet smaller efficiency for the same reason that
the reverse sinusoidal motions are less efficient than the forward
sinusoidal motions, in the absence of curls. The plusses, labeled ``Other,'' are for
other local optima, which have a variety of motions. Panel c shows
snapshots of a curling (but not self-intersecting) motion
in which the snake moves horizontally
leftward. The horizontal displacement is increased by 8/9 of a snake length between
snapshots to make them distinguishable. Panel d shows a different
curling motion in which the snake moves horizontally
rightward, and the snapshots are displaced vertically
to make them distinguishable. These alternative minima increase greatly
in number as $\mu_t$ decreases below 6, and some of them outperform
the forward sinusoidal motions at these transitional $\mu_t$ values.
Before discussing the regime of $\mu_t$ below 6,
we describe some scaling
properties for the traveling-wave solution to the
problem in the large-$\mu_t$ regime. In a separate paper
\cite{AlbenSnake2013II},
we derive analytically the form of the
optimal prescribed curvature in the limit of large $\mu_t$.
In that paper, we calculate that the optimal
curvature function is a periodic traveling wave with any functional form,
but in the limit that the
wavelength of the traveling wave tends to 0, and
such that the snake deflection has a root-mean-square amplitude of $2^{1/4}\mu_t^{-1/4}$.
The numerically-determined optima in figure \ref{fig:HighMutFig} do not
have small wavelengths, however, due to the finite number of modes used. This
is explained further in Supplementary Material
section \ref{sec:largemut}.

\begin{figure} [h]
           \begin{center}
           \begin{tabular}{c}
               \includegraphics[width=6in]{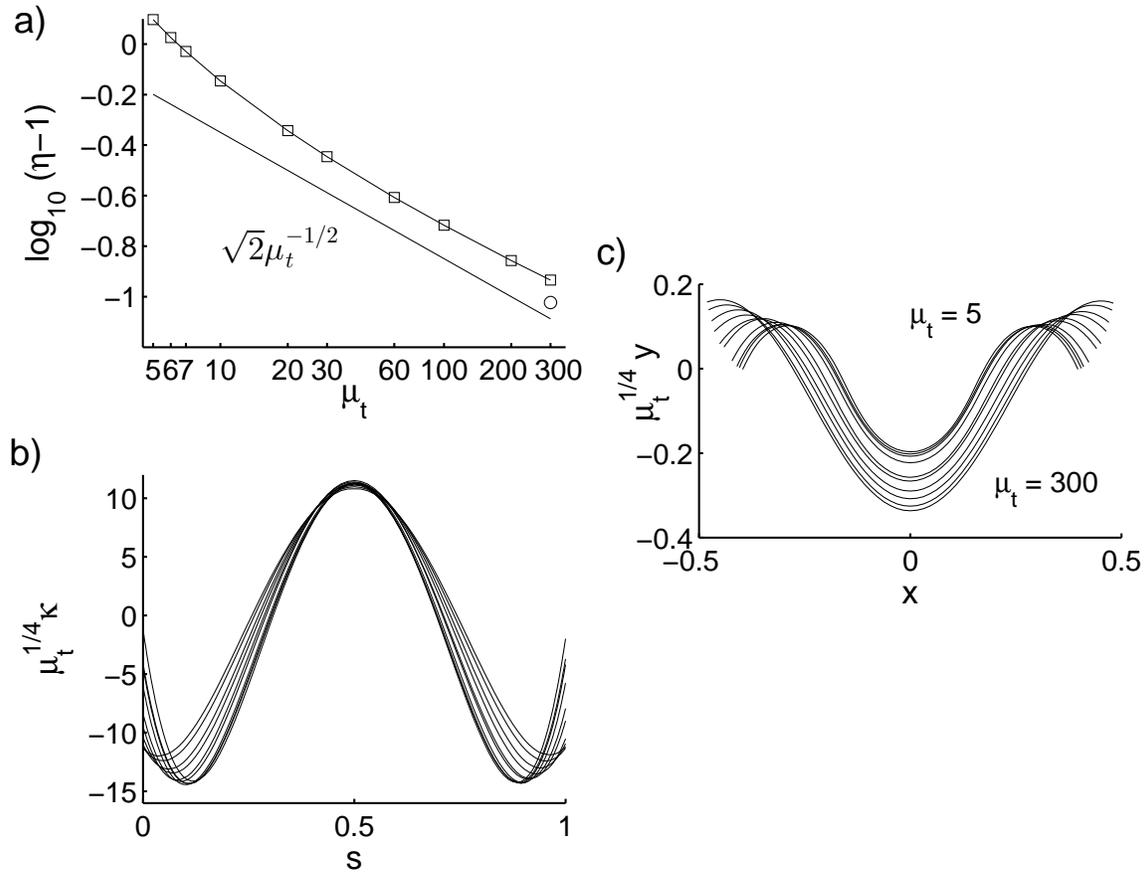} \\
           \vspace{-.25in}
           \end{tabular}
          \caption{\footnotesize The $\mu_t$-scalings of
the numerical traveling-wave optima
from figure \ref{fig:HighMutFig}. (a) The cost of locomotion $\eta$ minus
unity, on a log scale, for the numerical optima (squares) of
figure \ref{fig:HighMutFig}a together with $\eta-1$ for the analytical
optimum (solid line). (b) The curvature from figure \ref{fig:HighMutFig}b
rescaled by $\mu_t^{1/4}$. (c) The deflection from
figure \ref{fig:HighMutFig}c rescaled by $\mu_t^{1/4}$.
 \label{fig:RescaledHighMutFig}}
           \end{center}
         \vspace{-.10in}
        \end{figure}

We compare the numerically-found optima to the theoretical
results of \cite{AlbenSnake2013II} in figure \ref{fig:RescaledHighMutFig}. Here we take
the plots of figure \ref{fig:HighMutFig} and transform them according
to the theory. Figure \ref{fig:RescaledHighMutFig}a transforms $F$ from
figure \ref{fig:HighMutFig}a into $\eta$ by the formula $\eta = -e^2 F^{-1}$,
since rotations over one period are negligible for the
traveling-wave solutions. In figure
\ref{fig:RescaledHighMutFig}a
we plot $\eta-1$ versus $\mu_t$ for the numerical solutions (squares)
from figure \ref{fig:HighMutFig}a, along
with the small-wavelength theoretical result  (solid line), which is
\bqe
\eta \to 1 + \sqrt{2}\mu_t^{-1/2} + O(\mu_t^{-1}) \label{opteta}.
\eqe
\nn  The numerics
follow the same scaling as the theory, but with a consistent upward shift.
Some degree of upward shift is expected from the finite-mode truncation, which
makes the numerical result underperform the analytical result. The uniformity
of the shift may be due to the shape consistently assumed by the numerical optima
for various $\mu_t$. For $\mu_t = 300$ we have also computed $\eta$ via the
dynamical simulation for a much shorter-wavelength
shape given by $\kappa(s,t) = A_1\sin(2\pi s/U_w + 2 \pi t)$ with wavelength $U_w = 1/8$
and $A_1$ chosen so that
the snake deflection has a root-mean-square amplitude of $2^{1/4}\mu_t^{-1/4}$. The
result is given by the open circle, which outperforms the optimum identified in
the truncated-mode space, as expected. The circle is 0.013 above the solid line.
This deviation is due in part to the neglect of
the next term in the asymptotic expansion, proportional to $\mu_t^{-1}$ or 1/300 here.
Numerically, it is difficult to simulate motions with $U_w$ much smaller than 1/8.
Figure \ref{fig:RescaledHighMutFig}b shows the curvature, rescaled by $\mu_t^{1/4}$ to
give a collapse according to \cite{AlbenSnake2013II}. The data collapse well compared
to the unscaled data in \ref{fig:HighMutFig}b. Figure \ref{fig:RescaledHighMutFig}c
shows the body shapes from \ref{fig:HighMutFig}c with the vertical coordinate rescaled, and the
shapes plotted with all centers of mass located at the origin. We again find a good collapse,
consistent with the curvature collapse.

\section{Moderate to small $\mu_t$: from retrograde waves to standing
waves to direct waves}

Near $\mu_t = 6$ there is a qualitative change in the numerically-computed optima.
At values of $\mu_t \gtrsim 6$, the global minimizers of $F$ all have
backward-moving traveling waves which propel the snake forward.
In addition, there are local minimizers in which the waves
move forward along the snake (and the snake moves backward), and overlapping
solutions with integer numbers of curls appearing in the traveling waves. The
values of $F$ at the local minima, shown in figure \ref{fig:HighMutReverseFig},
are generally well-separated. In addition, there are a small number of local
minima with ratcheting types of motions. For $\mu_t \leq 6$ in
figure \ref{fig:HighMutReverseFig}, there are an increasing number of local
minimizers, denoted ``Other,''  quite close to the global minimizer. These
motions are perturbed traveling waves. The number of local minima
increases greatly as $\mu_t$ decreases below 6, and their
$F$ values are no longer well-separated.

\begin{figure} [h]
           \begin{center}
           \begin{tabular}{c}
               \includegraphics[width=4.5in]{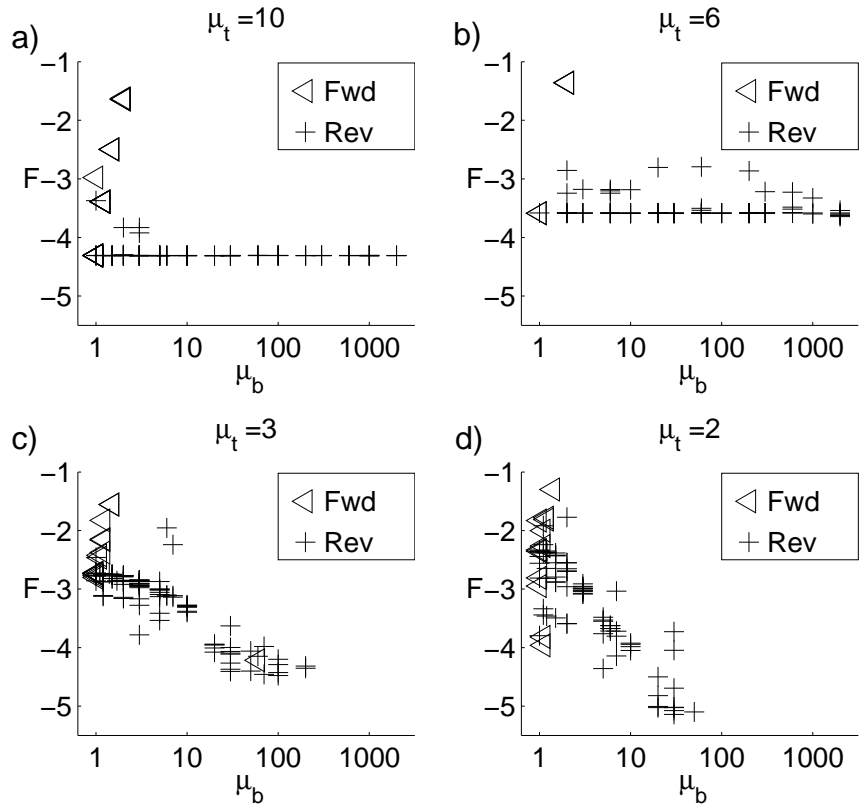} \\
           \vspace{-.25in}
           \end{tabular}
          \caption{\footnotesize Values of $F$ for all computed
local optima with various random initial iterates, across
$\mu_b$ at four $\mu_t$ (10 (a), 6 (b), 3 (c) and 2 (d))
corresponding to the emergence of perturbed-traveling-wave optima.
 \label{fig:TransitionalDataFig}}
           \end{center}
         \vspace{-.10in}
        \end{figure}

Figure \ref{fig:TransitionalDataFig} shows the values of $F$
for all computed local optima at four values of $\mu_t$
near the transition: 2, 3, 6, and 10. For each $\mu_t$, the
$F$ values are plotted against $\mu_b$. Also,
symbols are used to distinguish whether the local maximum in curvature
travels forward or backward on average, and hence gives the direction
of the curvature traveling wave (or perturbed traveling wave).
At $\mu_t = 10$ we find that the smallest $F$ is nearly constant
with respect to $\mu_b$. The same is true for $\mu_t = 6$
except at the largest $\mu_b = 2000$, for which there are several closely
spaced values of $F$ which are slightly below the minimal $F$
at smaller $\mu_b$. These values mark the first appearance of
the perturbed traveling waves as $\mu_t$ decreases. For $\mu_t = 3$,
the distribution of values is quite different, with a much stronger
decrease in $F$ as $\mu_b$ increases. There are many clusters of
closely-spaced but distinct values of $F$, showing a large
increase in the number of local equilibria. At each $\mu_b$ the smallest
values of $F$ are spread over a band of about 0.5 in width. At
$\mu_t = 2$, the same qualitative trend holds, except that the decrease
in $F$ with $\mu_b$ is much greater. There is more scatter of values overall
and of those representing forward-moving traveling waves.
There is randomness in the sample of $F$ values,
particularly evident for $\mu_t \leq 3$, because they are values
for a finite sample of local minimizers starting from random initial $\kappa$
coefficients.

Intuitively, the minima for $\mu_t = 3$ represent perturbations of
traveling wave motions towards ``ratcheting''-type motions which
involve small amounts of backward motion in a portion of the snake to ``push''
the rest of the snake forward. The appearance of backward motion
explains the variation of $F$ with $\mu_b$. As $\mu_t$ decreases from
2000 to 6, $F$ for the traveling waves increase, and for
$\mu_t = 3$, other motions outperform the traveling wave motions
at sufficiently large $\mu_b$. For $\mu_t = 2$, the motions
tend more toward ratcheting-type motions, and we will investigate this
further subsequently.

Another way of viewing the transition is to
consider the large-$\mu_t$ solution as $\mu_t$ decreases.
In the snake motion of figure \ref{fig:FirstSinusoidalFig}c,
the snake ``slips'' tranversely to itself as it slithers
forward (rightward). This transverse motion has a component in the
backward (leftward) direction, and the resulting friction
with the ground propels the snake forward as a
thrust force from transverse friction, which balances the drag force from
the forward component of tangential friction.
The thrust force is proportional to $\mu_t$, and
increases as the slope of the snake increases,
which increases the component of transverse
friction in the horizontal direction. If $\mu_t$
is decreased, the thrust force can be maintained
by increasing the slope of the snake. This is
indeed observed in the sinusoidal optima as
$\mu_t$ decreases, in figure \ref{fig:HighMutFig}c.
When $\mu_t$ decreases to 3, the slope
of the snake becomes nearly vertical at its maximum,
and no further decreases in $\mu_t$ can be accommodated
by increasing the slope of the snake. Thus, near
this value of $\mu_t$, motions different from the
sinusoidal traveling wave become preferred.

% A given
% traveling wave motion will slip backward more as $\mu_t$ decreases,
% unless the amplitude increases. However, the component of the transverse
% force in the forward direction is maximum when the slope of the
% snake becomes vertical. Thus there is a limit to
% how much the slip can be decreased by increasing the amplitude of
% the traveling wave. When $\mu_t$ decreases to 3, the slope
% of the snake becomes nearly vertical at its maximum. In the
% traveling wave motion, The
% thrust provided by transverse friction is proportional to
% $\mu_t$ and the forward component of

\begin{figure} [h]
           \begin{center}
           \begin{tabular}{c}
               \includegraphics[width=6in]{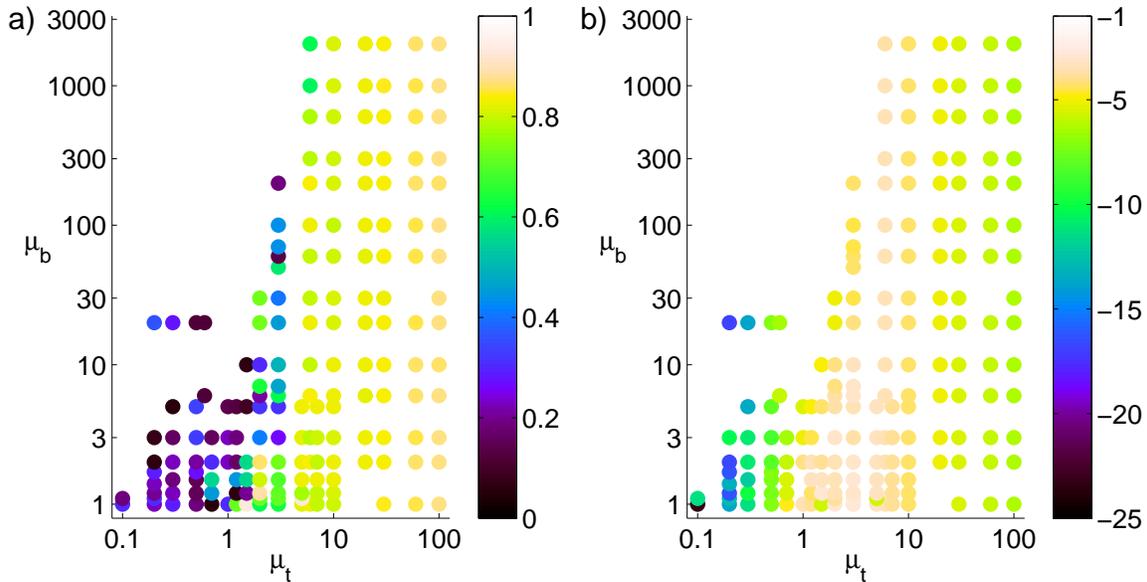} \\
           \vspace{-.25in}
           \end{tabular}
          \caption{\footnotesize (a) At each $(\mu_t, \mu_b)$,
the maximum value of the function $\psi$ over all
optima is plotted as a shaded dot. (b) The values of $F$
corresponding to the optima in (a) are plotted.
 \label{fig:PsiFig}}
           \end{center}
         \vspace{-.10in}
        \end{figure}

In figure \ref{fig:PsiFig}a we plot a function of the shape dynamics which
quantifies the transition away from traveling wave motions near $\mu_t = 6$.
The quantity is
\bqe
\psi[\kappa] \equiv \frac{\min_t \max_{0.2 \leq s \leq 0.8}{|\kappa(s,t)|}}
{\max_t \max_{0.2 \leq s \leq 0.8}{|\kappa(s,t)|}}.
\eqe
\nn For $\kappa(s,t)$ which is a traveling wave with a wavelength sufficiently
small that a maximum or minimum in curvature occurs
in $0.2 \leq s \leq 0.8$ at all times, and for which the maximum
and minimum are equal in magnitude, $\psi$ equals unity. For $\mu_t \gtrsim 6$,
the optimal $\kappa(s,t)$ are approximately such traveling waves; we restrict
$s$ away from the ends to avoid slight intensifications in curvatures near the ends.
In figure \ref{fig:PsiFig}a we plot the maximum of $\psi$ over all
equilibria located at a given $(\mu_t, \mu_b)$ pair.

We find $\psi > 0.8$ for $\mu_t \gtrsim 6$. $\psi$ increases
slightly from about 0.8 to 0.9 as $\mu_t$ increases from 6 to 200.
For $\mu_t$ = 6, $\psi$ decreases slightly but noticeably, from
0.8 to 0.7, as $\mu_b$ exceeds 200, showing that at
transitional $\mu_t$, the perturbations
away from the traveling wave solutions appear first at the largest
$\mu_b$. At $\mu_t$ = 3, the values of $\psi$ have decreased to
around 0.6 for most $\mu_b$; $\psi$ is as large as 0.9
at smaller $\mu_b$ and as small as 0.4 at larger $\mu_b$. For
$\mu_t < 3$, the values of $\psi$ are generally much smaller.
There is considerable randomness in the values of $\psi$ plotted for
$\mu_t < 3$ because they are the maxima over samples
of local equilibria starting from 10-20 random initial $\kappa$
coefficient vectors.

In figure \ref{fig:PsiFig}b we plot the values of $F$ for
the optima whose $\psi$-values are shown in panel a. We find
that $F$ varies smoothly over the region of transitional
$\mu_t$, $2 \leq \mu_t \leq 10$,
attaining its highest values there. The retrograde
traveling-wave motions and other motions are similarly efficient
at a transitional $\mu_t$, and the other motions
become preferable as $\mu_t$ decreases.

\begin{figure} [h]
           \begin{center}
           \begin{tabular}{c}
               \includegraphics[width=5in]{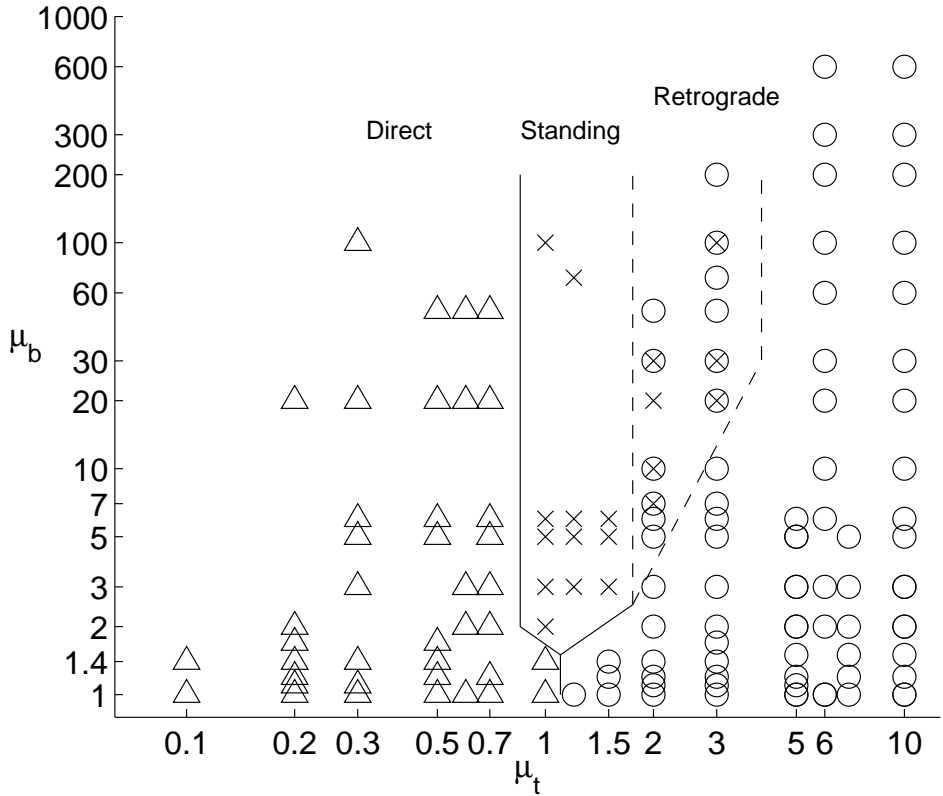} \\
           \vspace{-.25in}
           \end{tabular}
          \caption{\footnotesize Phase diagram of optimal motions
in $(\mu_t, \mu_b)$-space.
 \label{fig:PhasePlaneFig}}
           \end{center}
         \vspace{-.10in}
        \end{figure}

A global picture of the optimal states is presented in the ``phase
diagram'' of figure \ref{fig:PhasePlaneFig}. At large $\mu_t$, the
retrograde traveling waves are optimal. At small $\mu_t$, all
of the optima have the form of direct traveling waves. Unlike
for the large-$\mu_t$ waves, these direct waves all have large
amplitudes, and their specific form varies with $\mu_b$. Near
$\mu_t = 1$ there is a region where standing waves with non-zero
mean deflection (or ratcheting motions) are seen.
Examples will be given shortly. These optima co-exist with
retrograde-wave optima at larger $\mu_t$. We note that
there are other optima which do not fit easily into one of
these categories, throughout the phase plane. These
other optima are generally found less often by the computational
search, and are
generally suboptimal to the traveling waves, but there
are exceptions. For the sake of brevity we do not
discuss such motions here.
%Also, near the
%transitions shown by dashed and solid lines in figure
% \ref{fig:PhasePlaneFig}, it can be difficult to clearly classify
%some of the optima.

To obtain a clear picture of the snake kinematics across
the transitional region shown in figure \ref{fig:PhasePlaneFig},
we give figures showing snapshots of snake kinematics,
together with spatiotemporal curvature plots, at $\mu_b$ =
1 and 3 in this section and at $\mu_b$ = 2, 5, and 20
in Supplementary Material section \ref{sec:modmut}. Each of these figures
represents one horizontal ``slice'' through
the phase diagram in figure \ref{fig:PhasePlaneFig}.

\begin{figure} [h]
           \begin{center}
           \begin{tabular}{c}
               \includegraphics[width=5in]{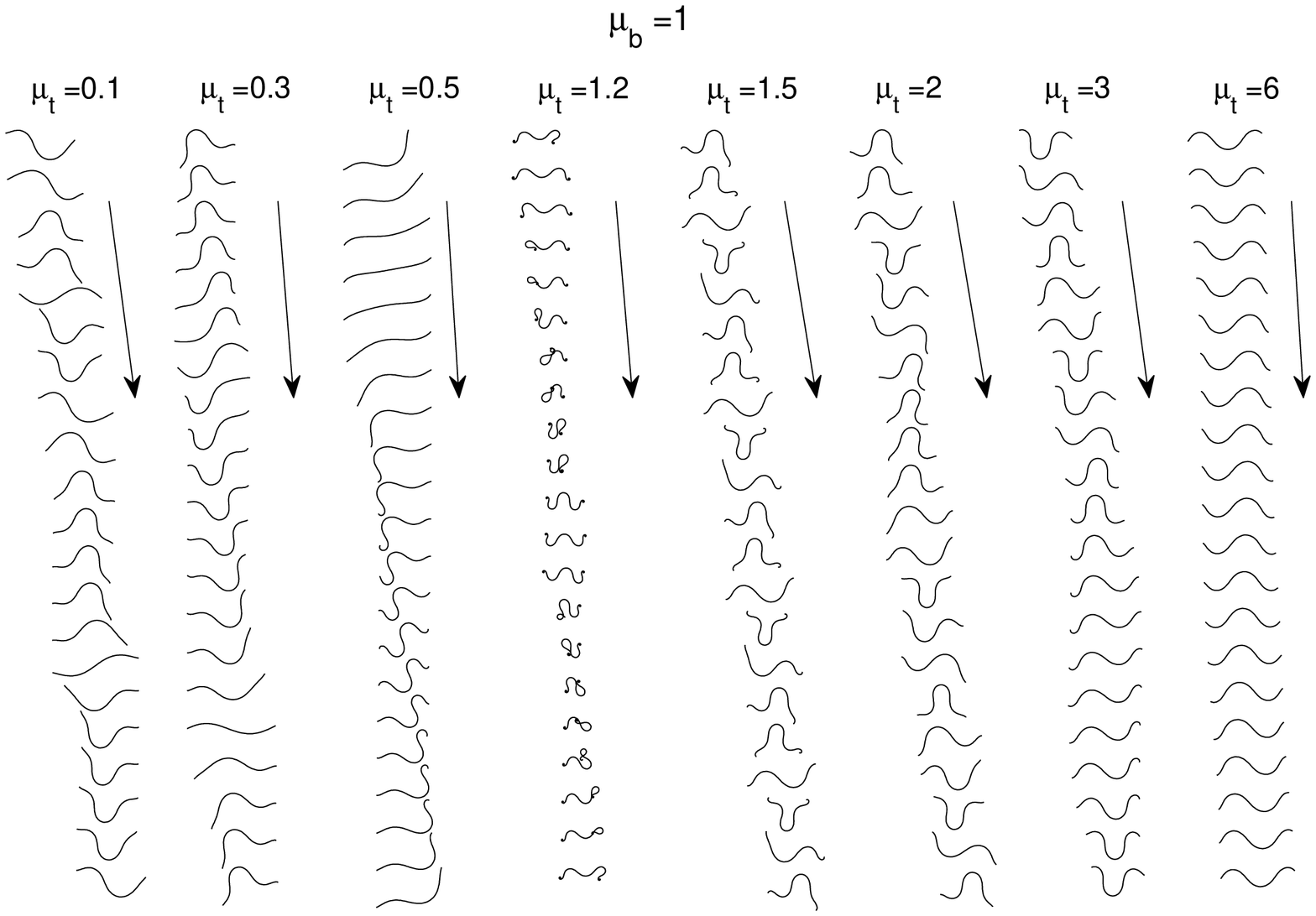} \\
           \vspace{-.15in}
\includegraphics[width=5in]{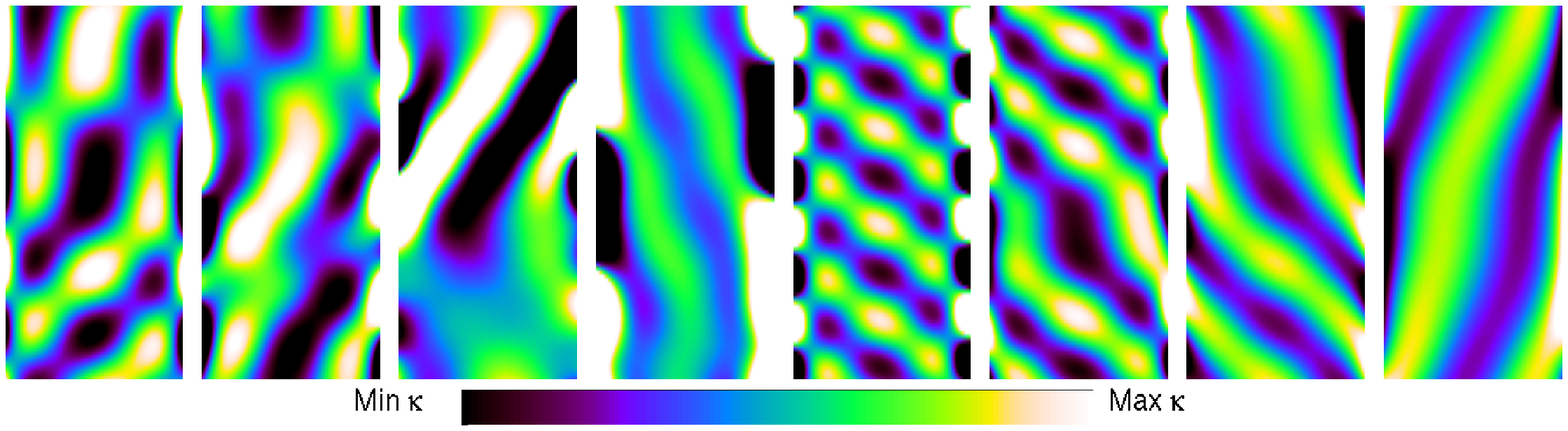} \\
           \end{tabular}
          \caption{\footnotesize Snapshots of optimal snake motions at
various $\mu_t$ (one per column, labeled at the top), for $\mu_b = 1$.
 \label{fig:Multi1}}
           \end{center}
         \vspace{-.0in}
        \end{figure}

In figure \ref{fig:Multi1}, we show the optimal snake
kinematics for $\mu_b = 1$ and eight $\mu_t$ spanning the
transition. Each set of snapshots is arranged in a single
column, moving downward as time moves forward over one period,
from top to bottom (shown by arrows). Similarly to
figure \ref{fig:HighMutReverseFig}b and d, the snapshots are shown in a frame in which
the net motion is purely horizontal, from left to right.
However, the snapshots are given an extra uniform
vertical spacing to make
them visable (i.e. to prevent overlapping).
Below the snapshots,
the corresponding curvature plot is shown, in the same format
as in figures \ref{fig:FirstSinusoidalFig} and
\ref{fig:AltSinusoidalFig}, with $s$ increasing from 0 to 1 on
the horizontal axis, from left to right, and $t$ increasing
from 0 to 1, from bottom to top. We have omitted the axes,
and the specific
curvature values for each plot, instead using a single color bar (shown
at bottom) with curvature values which range from the maximum to
the minimum for each plot. By omitting the axes we are
able to show several of the optimal motions side-by-side. By comparing
a large number of motions in a single figure, it is easier to see the
general trends.

We begin by discussing the rightmost column in figure \ref{fig:Multi1},
with $\mu_t = 6$. A retrograde traveling wave is clearly seen, so this
optimum is a continuation of the large-$\mu_t$ solutions. At this
$\mu_t$, just above the transition, the wave amplitude is large,
and the wave clearly moves backwards in the lab frame. Thus the snake
slips backward considerably even though it moves forward, because
$\mu_t$ is not very large. Moving one column to the left,
the optimum at $\mu_t = 3$ shows a clear difference. At certain
instants, the curvature becomes locally intensified near the
snake midpoint. This occurs when the local curvature maximum reaches
the snake midpoint. The motion is still that of a retrograde
traveling wave, but now the wave shape changes over the period. Hence
the displacement is not a traveling wave
of the form $g(s+U_w t)$ for
a fixed periodic function $g$. Comparing
the curvature plots for $\mu_t = 6$ and 3, we see that the bands
have nearly constant width across time for $\mu_t = 6$. At
$\mu_t = 3$, by contrast, the curvature bands become modulated.
The bands show bulges near the curvature intensifications. Moving
leftward again, to $\mu_t = 2$ and then to 1.5, the
curvature intensifications increase further, showing
a general trend. In all cases, however, the curvature peaks
move backward along the snake, even as the peak values change,
so these motions are retrograde waves. At $\mu_t = 1.2$, a novel
feature is seen: the endpoints of the snake curl up (but do not
self-intersect), forming ``anchor points.'' Between these points,
the remainder of the snake moves forward. In the curvature plot, bands
can still be seen corresponding to retrograde waves. The three remaining
columns, moving leftward, all show direct waves. That is, the wave moves
in the same direction as the snake \cite{lauga2006tuning}.
No optimum is shown for $\mu_t = 1$.
The point $\mu_b = 1, \mu_t = 1$, in which friction is equal in all
directions, is in some sense a singular point for our computations.
Near this point, it is difficult to obtain convergence to
optimal motions, and optima are
difficult to classify clearly as retrograde or direct waves (or another
simple motion). Nonetheless,
we have computed one optimum which can be classifed as a direct
wave, but it has complicated and special features which are not easy to
classify. In the three left columns, the direct waves can be recognized
by the bands in the curvature
plots. The bands have the direction of direct waves.  However,
the bands are not nearly as simple in form as those for
the retrograde waves, at large $\mu_t$.

\begin{figure} [h]
           \begin{center}
           \begin{tabular}{c}
               \includegraphics[width=5in]{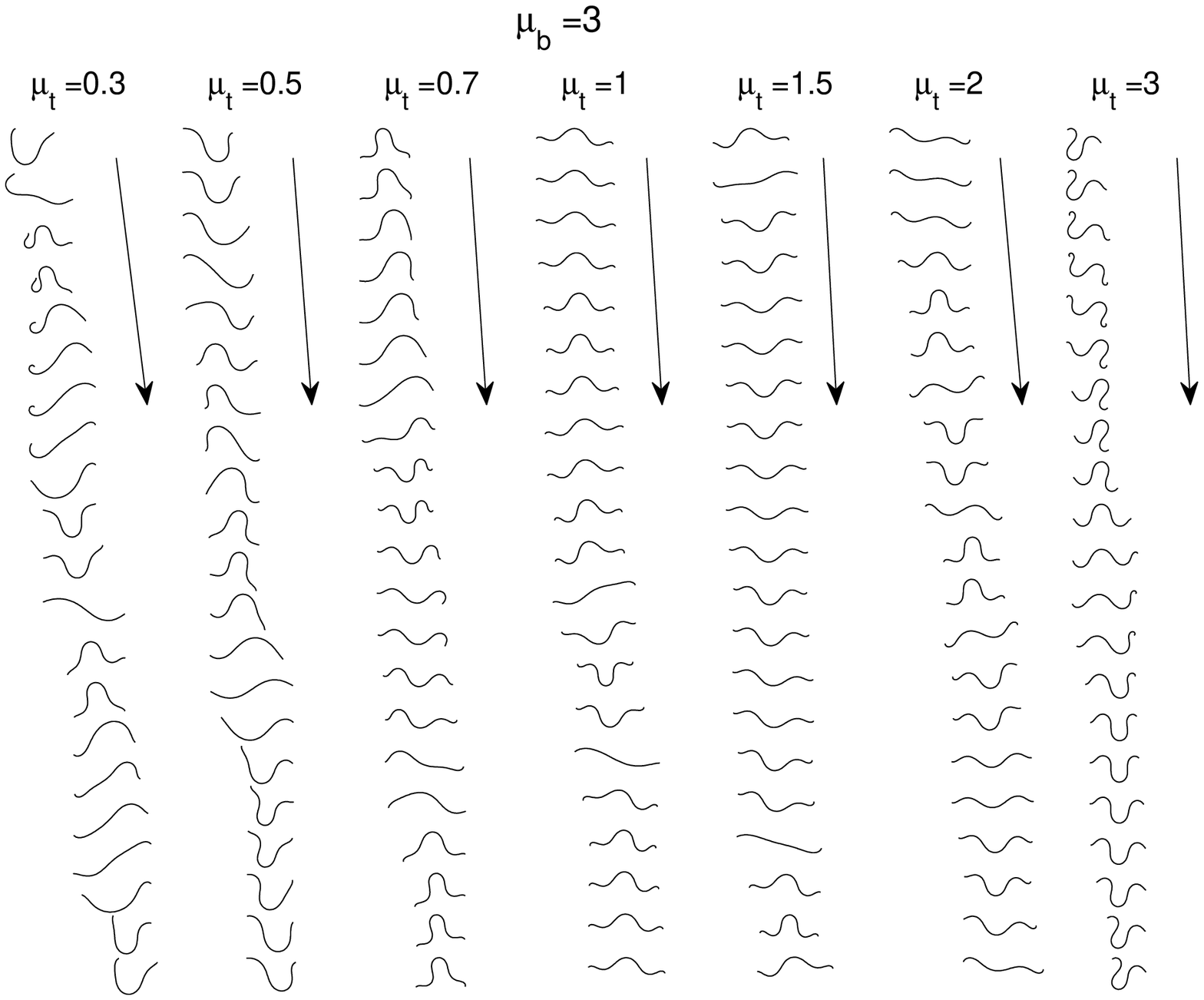} \\
           \vspace{-.15in}
\includegraphics[width=5in]{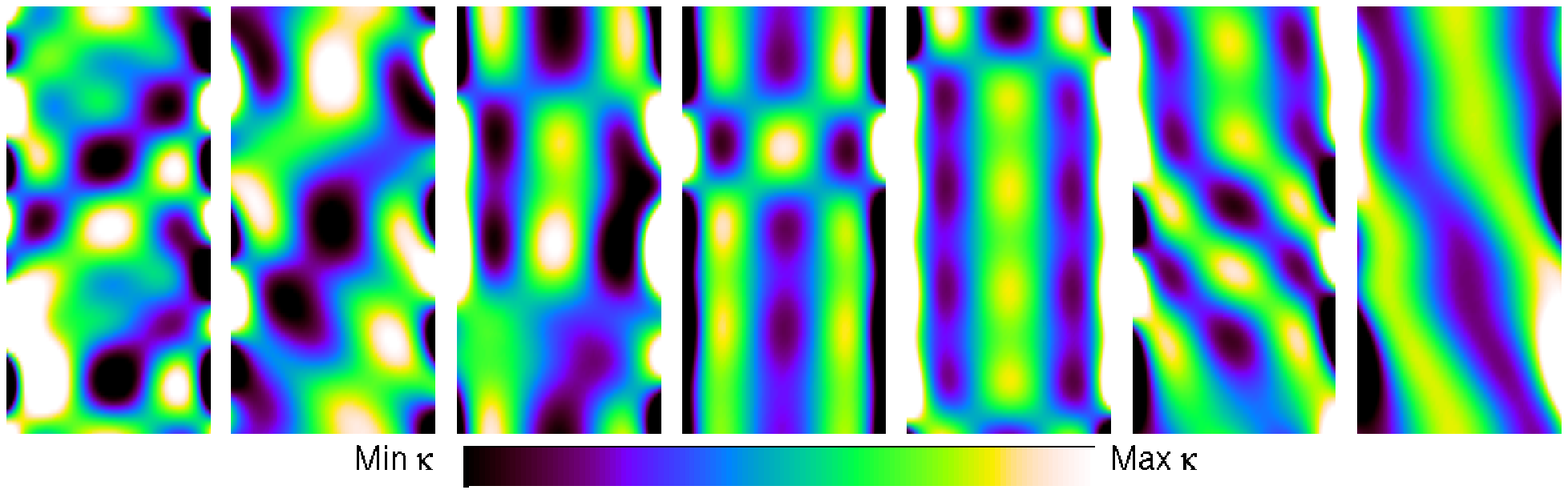} \\
           \end{tabular}
          \caption{\footnotesize Snapshots of optimal snake motions at
various $\mu_t$ (one per column, labeled at the top), for $\mu_b = 3$.
 \label{fig:Multi3}}
           \end{center}
         \vspace{-.0in}
        \end{figure}

Figure \ref{fig:Multi3} shows optima for
$\mu_b$ now increased to 3. The same transition is seen, but now
in two middle columns, with $\mu_t = 1$ and 1.5, standing waves
are seen. The curvature plots are nearly left-right symmetric.
In the corresponding motions, the snake bends and unbends itself.
Because $\mu_b > 1$, there is a preference for motion in the forward
direction even though the curvature is nearly symmetric with respect
to $s = 1/2$. These optima did not occur for $\mu_b = 1$
(figure \ref{fig:Multi1}) because symmetric motions with
equal forward and backward friction
would yield zero net distance traveled. It is somewhat
surprising that such motions can be completely ineffective
for $\mu_b = 1$ and yet represent optima for $\mu_b$ as small
as 2, as shown in figure \ref{fig:PhasePlaneFig}.

%\section{Moderate to small $\mu_t$: some other interesting minima}

%transverse motion
%anchor and swing
We have presented only a limited selection of optima for
moderate $\mu_t$, and largely with
pictures, because it is difficult to describe and classify many
of the motions we have observed. Future work may consider this
interesting region of parameter space further.

We now move on to the limit of zero $\mu_t$. In this limit,
we find a direct-wave motion with zero cost of locomotion. This
explains the preference for direct waves at smaller $\mu_t$
in this section.

\section{Zero $\mu_t$: direct-wave locomotion at zero cost}

When $\mu_t = 0$, it is possible to define a simple shape dynamics with zero
cost of locomotion. The shape moves with a nonzero speed while doing zero
work. It is a traveling wave with the shape of a triangle wave which has zero mean
vertical deflection:
\bqe
y(s,t) = A_0 \barint \mbox{sgn}(\sin(2\pi (s-t))) ds. \label{y}
\eqe
\nn The vertical speed corresponding to (\ref{y}) is a square wave:
\bqe
\partial_t y(s,t) = -A_0 \mbox{sgn}(\sin(2\pi (s-t))). \label{dty}
\eqe
The wave of vertical deflection moves forward with constant speed one,
and the  horizontal motion is also forward, with a constant speed $U$:
\bqe
\partial_t x(s,t) = U \label{dtx}
\eqe
\nn We determine $U$ by setting the net horizontal force on the snake to zero:
\bqe
\int  \widehat{\partial_t{\mathbf{X}}}\cdot \hat{\mathbf{s}} \,s_x\, ds = 0.
\label{fxalt}
\eqe
\nn with
\bqe
\hat{\mathbf{s}} = \left( {\begin{array}{c} \sqrt{1-A_0^2} \\ A_0 \mbox{sgn}(\sin(2\pi (s-t))) \end{array} } \right) \label{stri}
\eqe
\nn Solving (\ref{fxalt}) for $U$ we obtain
\bqe
U = \frac{A_0^2}{\sqrt{1-A_0^2}}. \label{Usol}
\eqe
\nn With this choice of $U$, $\widehat{\partial_t{\mathbf{X}}}\cdot \hat{\mathbf{s}}$
is identically zero on the snake, so the motion is purely in the normal
direction. Therefore, there are no forces and torques on the snake, and
no work done against friction. However, the snake moves forward at a nonzero speed $U$. This solution has infinite curvature at the peaks and troughs of the
triangle wave, and we may ask if zero cost of locomotion ($\eta$) can be obtained
in the limit of a sequence of smooth curvature functions which tend to this
singular curvature function. To answer this question, we have set $y$ to
the Fourier series approximation to (\ref{y}) with $n$ wave numbers,
for $n$ ranging from 2 to 1000. At each time $t$, we have then calculated a
net horizontal speed $U(t)$, net vertical speed $V(t)$ and a net
rotational speed $dR(t)/dt$ such that $F_x(t) = 0$,
$F_y(t) = 0$ and $\tau(t) = 0$. We find very clear asymptotic behaviors
for $U(t)$, $V(t)$, $dR(t)/dt$, and $\eta$:
\bqe
U(t) = \frac{A_0^2}{\sqrt{1-A_0^2}} + O(n^{-1})\;,\; V(t) = O(n^{-1})\;, \;
\frac{dR}{dt} = O(n^{-1})\;, \;\eta = O(n^{-1}).
\eqe
\nn Therefore
zero cost of locomotion does in fact arise in the limit of a
sequence of smooth shape dynamics. This type of traveling wave
shape dynamics, which yields locomotion in the
direction of the wave propogation,
has been called a ``direct'' wave
in the context of crawling by snails \citep{lauga2006tuning}
and worms \citep{sfakiotakis2009undulatory}.

The triangular traveling wave motion is particularly interesting because it
can represent an optimal motion at both zero and infinite $\mu_t$, and
its motion can be computed analytically for all $\mu_t$. We
show how this motion embodies many aspects of
the general problem in Supplementary Material section
\ref{sec:trimut}.

\section{Conclusion \label{sec:Concl}}

We have studied the optimization of planar
snake motions for efficiency, using a fairly simple model
for the motion of snakes using friction. Our numerical optimization
is performed in a space of limited dimension (45), but which is
large enough to allow for a wide range of motions.
The optimal
motions have a clear pattern when the coefficient of transverse friction is
large. Our numerical results
indicate that a traveling-wave motion of small amplitude
is optimal, which agrees with an asymptotic analysis in
\cite{AlbenSnake2013II}. When this coefficient is small, another
traveling-wave motion, of large amplitude, is optimal.
When the coefficient
of transverse friction is neither large nor small, the optimal
motions are far from simple, and we have only begun to address
this regime.

A natural extension of this work is to include three-dimensional
motions of snakes. In other words, one would give the snake the ability
to lift itself off of the plane \citep{GuMa2008a,HuNiScSh2009a,HuSh2012a},
as occurs in many familiar motions such as side-winding \citep{jayne1986kinematics}.
Another interesting direction is to consider the motions of snakes
in the presence of confining walls or barriers
\citep{MaHu2012a,majmudar2012experiments}.

\begin{acknowledgments}
We would like to acknowledge helpful discussions on snake
physiology and mechanics with David Hu and Hamidreza Marvi,
helpful discussions with Fangxu Jing during our previous
study of two- and three-link snakes, and
the support of NSF-DMS
Mathematical Biology Grant 1022619 and a Sloan Research Fellowship.
\end{acknowledgments}

\bibliographystyle{unsrt}
\bibliography{snake}

\newpage

\noindent
{\large Supplementary material for:\\ Optimizing snake locomotion in the plane. I. Computations}
\appendix
\noindent\section{Supplementary Material: Objective function \label{sec:F}}
\setcounter{page}{1}
To understand our choice of $F$,
we first describe the class
of motions $\mathbf{X}(s,t)$ which result from periodic-in-time
$\kappa(s,t)$. Let $\mathbf{X}(s,t)$ and $\partial_t\mathbf{X}(s,t)$
be given for the moment. Let us rotate $\mathbf{X}(s,t)$ by an angle
$\alpha$ and translate $\mathbf{X}(s,t)$ by a vector $\mathbf{v}$,
uniformly over $s$, so we have a rigid body motion. Then $\hat{\mathbf{s}}$ and $\hat{\mathbf{n}}$
are also uniformly rotated by $\alpha$. Let us
also rotate $\partial_t\mathbf{X}(s,t)$ uniformly by $\alpha$.
Then by (\ref{friction1}), $\mathbf{f}$ is uniformly rotated by $\alpha$.
Hence, if $\mathbf{X}(s,t)$ and $\partial_t\mathbf{X}(s,t)$ are such
that the force and torque balances (\ref{fxb})--(\ref{torqueb}) hold,
they will continue to hold if we apply a rigid-body rotation and
translation to $\mathbf{X}(s,t)$ and rotate $\partial_t\mathbf{X}(s,t)$
by the same angle. Now, since $\kappa(s,t+1) = \kappa(s,t)$,
$\mathbf{X}(s,t+1)$ is obtained from $\mathbf{X}(s,t)$ by a rigid-body
rotation by an angle $\alpha_1$ and translation by a vector $\mathbf{v}_1$.
In fact, the trajectory of the body during each period is the same as during
the previous period but with a rigid translation by $\mathbf{v}_1$ and
rotation by $\alpha_1$.
Over many periods such a body follows a circular path when $\alpha_1 \neq 0$,
and because of the rotation,
the distance traveled over $n$ periods may be much smaller than $n$ times
the distance traveled over one period. To obtain a motion which yields
a large distance over many periods, we restrict
to body motions for which $\mathbf{X}(s,t+1)$ is obtained from $\mathbf{X}(s,t)$
by a translation only, with rotation angle $\alpha_1 = 0$. We
implement this constraint approximately
in our optimization problem, using a penalization term.
We multiply $d$ in (\ref{eta}) by a factor which penalizes rotations over
a single period:
\bqe
d \rightarrow d e^{2\cos(\theta(0,1)-\theta(0,0))} \label{penalty}
\eqe
\nn
(\ref{penalty}) involves the change in angle at the snake's trailing edge over a
period, which is the same as $\alpha_1$.
The exponential factor has a maximum of $e^2$ when $\alpha_1 = 0$, and
a minimum of $e^{-2}$ when $\alpha_1 = \pi$.
Using the factor of 2 in the exponent
of (\ref{penalty}), all of the optima found by our computations
involve very little rotation---less
than 0.5 degrees over all of $(\mu_b, \mu_t)$ space.
When the factor is decreased from 2, some of the optima begin to involve
nontrivial amounts of rotation. In short, the factor of 2 represents
a choice of how much to weight lack of
rotation relative to distance and work in the objective function.
As described in the results section,
for $\mu_t \gtrsim 6$ and all $\mu_b$, we find optima which have an
analytical traveling wave form, and in the limit of large $\mu_t$,
rotations tend to zero for these solutions. These optima are found
even with no rotation penalty, so the rotation penalty plays a negligible
role in this region of parameter space. For smaller $\mu_t$, however, in the
absence of the rotation penalty, many of
the optima have significant rotations.

A second modification to $\eta$ is made to avoid large values of
$\eta$. There are snake motions (i.e. with the $\kappa(s,t) = \kappa(1-s,t)$ symmetry and
$\mu_b = 1$) which have $d = 0$ and $W \neq 0$, giving $\eta = \infty$. The minimization
search commonly encounters motions which are nearly as ineffective
throughout shape space, and $\eta$ has a large gradient in the vicinity
of such motions.
%Both $\eta$ and its gradient are large
%, because
%these motions occur , and they are close to
%effective motions (in terms of $\kappa \in L^2([0,1]\times[0,1])$). Thus
%$\eta$ is large and rapidly fluctuating in the space of
%$\kappa \in L^2([0,1]\times[0,1])$, and
It is challenging to
minimize such a function numerically.
We therefore substitute
$-d/W$ for $W/d$. The minima are the same for both because each is an increasing
function of the other, but $-d/W$ varies much more smoothly in most of
shape parameter space.
There are cases where $-d/W$ tends to $-\infty$, but these
only occur in a small region of parameter space which is rarely approached
by our computations.

Combining these ideas, the function we minimize in place of $\eta$ is
\bqe
F = -\frac{d}{W} e^{2\cos(\theta(0,1)-\theta(0,0))}. \label{Fa}
\eqe

\noindent\section{Supplementary Material:
Invariance under reparametrization of time \label{sec:t}}

Here we show that $d$, $W$, and $F$ are the same for any reparametrization of
time---in other words, for $\kappa(s,\gamma(t))$ with any nondecreasing differentiable mapping
$\gamma$ from $[0,1]$ to $[0,1]$ with $\gamma(0) = 0$ and
$\gamma(1) = 1$. Such a reparametrization can be used to
reduce the high-frequency components of a motion while keeping the efficiency
the same. Therefore, it is reasonable to expect that a good approximation to
any minimizer of $F$ can be obtained with low
temporal frequencies, that is, with an expansion in the form of (\ref{expn}) with
small $m_1$. To show that $d$, $W$, and $F$ are the
same for a reparametrization of time, let $\kappa(s,t)$ be given and
$\gamma(t)$ be the aforementioned parametrization function. Let $\mathbf{X}(s,t)$
be the motion obtained by integrating $\kappa(s,t)$ from $s = 0$ as in
(\ref{theta0})--(\ref{y0})
with the integration
constants $\mathbf{X}_0(t) = \mathbf{X}(0,t)$ and $\theta_0(t)$ chosen
to satisfy the dynamical equations (\ref{fxb})--(\ref{torqueb}).
Now we show that $\tilde{\mathbf{X}}(s,t) = \mathbf{X}(s,\gamma(t))$ also satisfies
(\ref{fxb})--(\ref{torqueb}) at each time $t$. By the chain rule
of differentiation,
\bqe
\widehat{\partial_t\tilde{\mathbf{X}}}(s,t) =
\frac{d\mathbf{X}(s,\gamma(t))/dt}{\|d\mathbf{X}(s,\gamma(t))/dt \|}
= \frac{\gamma'(t)\partial_\gamma{\mathbf{X}}(s,\gamma)|_{\gamma = \gamma(t)}}
{\|\gamma'(t)\partial_\gamma{\mathbf{X}}(s,\gamma)|_{\gamma = \gamma(t)} \|}
= \widehat{\partial_\gamma \mathbf{X}}(s,\gamma)|_{\gamma = \gamma(t)} \label{paramvel}
\eqe
\nn using $\gamma'(t) \geq 0$. Let $\tilde{\mathbf{f}}(s,t)$ be
the force distribution corresponding to
$\tilde{\mathbf{X}}(s,\gamma(t))$. It is given by (\ref{friction1})
with $\widehat{\partial_t \tilde{\mathbf{X}}}$ in place of
$\widehat{\partial_t \mathbf{X}}$ and
$(\tilde{\hat{\mathbf{s}}}, \tilde{\hat{\mathbf{n}}})$ in place of
$(\hat{\mathbf{s}}, \hat{\mathbf{n}})$, where $(\tilde{\hat{\mathbf{s}}}, \tilde{\hat{\mathbf{n}}})$
are the unit vectors tangent and normal to
$\tilde{\mathbf{X}}(s,t)$. Using (\ref{paramvel})
we obtain $\tilde{\mathbf{f}}(s,t) = \mathbf{f}(s,\gamma(t))$.
Since (\ref{fxb})--(\ref{torqueb}) hold for
$\mathbf{f}(s,t)$ for all $t \in [0,1]$, they also hold for
$\mathbf{f}(s,\gamma(t))$ for all $t \in [0,1]$
since $\gamma(t) \in [0,1]$. Thus they hold for
$\tilde{\mathbf{f}}(s,t)$ in place of $\mathbf{f}(s,t)$,
so $\tilde{\mathbf{X}}(s,t)$ satisfies the dynamical equations
(\ref{fxb})--(\ref{torqueb}). The distance traveled
$\tilde{d}$ for this motion is
\begin{align}
\tilde{d} &= \sqrt{\left(\int_0^1 \int_0^1 \partial_t \tilde{x}(s,t) ds dt\right)^2 +
\left(\int_0^1 \int_0^1 \partial_t \tilde{y}(s,t)ds dt\right)^2}. \label{dista} \\
&= \sqrt{\left(\int_0^1 \int_0^1 \partial_\gamma x(s,\gamma) ds d\gamma\right)^2 +
\left(\int_0^1 \int_0^1 \partial_\gamma y(s,\gamma)ds d\gamma\right)^2} \\
&= d
\end{align}
\nn using (\ref{dist}). The work done is
\begin{align}
\tilde{W} &= \int_0^1 \int_0^1 \tilde{\mathbf{f}}(s,t)
\cdot \partial_t\tilde{\mathbf{X}}(s,t) ds dt \label{Wa}
= \int_0^1 \int_0^1 \mathbf{f}(s,\gamma)
\cdot \partial_\gamma\mathbf{X}(s,\gamma) ds d\gamma
= W
\end{align}
\nn using (\ref{W}). The net rotation over a period is
\begin{align}
\tilde{\theta}(0,1) - \tilde{\theta}(0,0) =
\int_0^1 \partial_t \tilde{\theta}(0,t) dt
= \int_0^1 \partial_\gamma \theta(0,\gamma) d\gamma
= \theta(0,1) - \theta(0,0),
\end{align}
\nn so $F$ (\ref{F}) is unchanged by the reparametrization of time.

\noindent\section{Supplementary Material:
Effect of finite mode truncation on large-$\mu_t$ optimum \label{sec:largemut}}

The theory of \cite{AlbenSnake2013II} predicts that the wavelengths of the optimal shapes
should tend to zero, but the wavelengths for
the computed optima shown in figure \ref{fig:HighMutFig}b and c are all in the
neighborhood of one body length, not particularly small. The reason is
that it is not possible to represent traveling waves with very small wavelengths
with the value of $n_1$ (5) used here, which includes polynomials in $s$
of up to fourth degree.

\begin{figure} [h]
           \begin{center}
           \begin{tabular}{c}
               \includegraphics[width=5in]{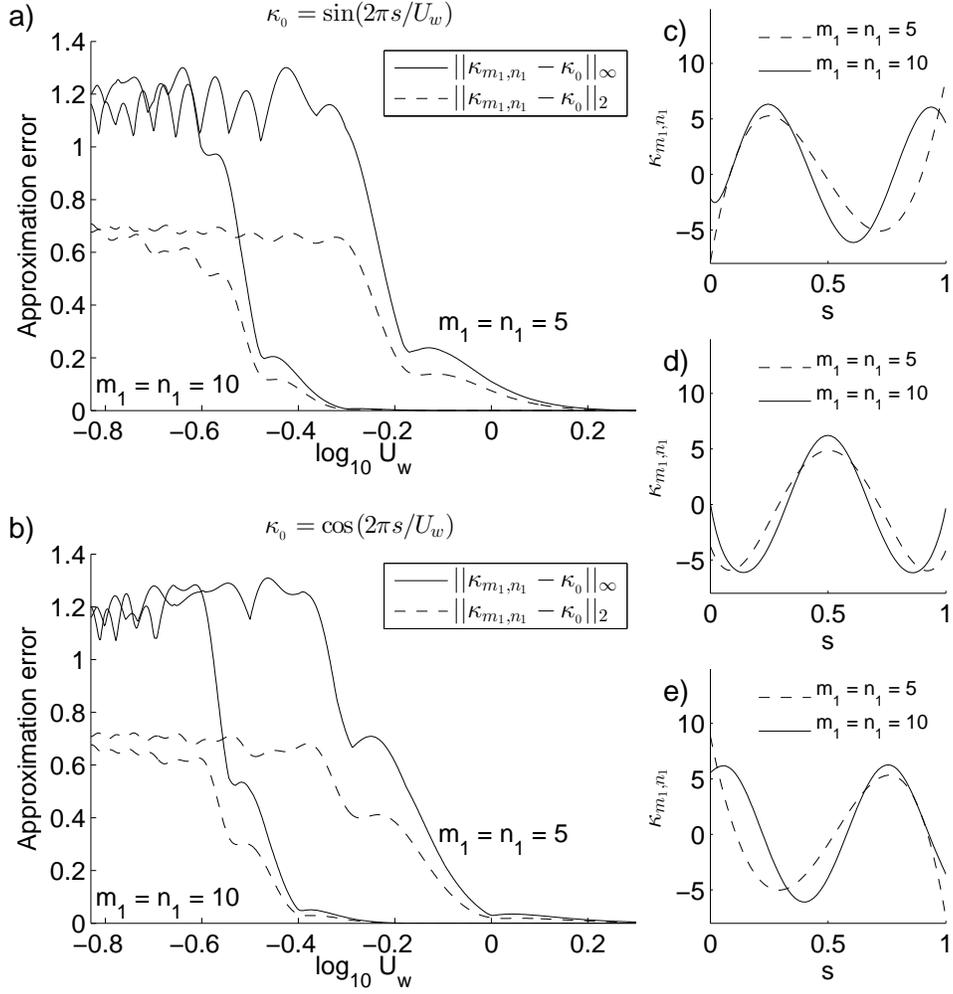} \\
           \vspace{-.25in}
           \end{tabular}
          \caption{\footnotesize The approximation error in representing
(a) sine and (b) cosine functions with different wavenumbers using
either five or ten Chebyshev modes (corresponding curves are labeled).
Panels c-e show snapshots of curvature versus arc length in the optimal
numerical traveling wave motions with either five (dashed line)
or ten (solid line) Fourier and Chebyshev modes. The snapshots
are at times when there are local curvature maxima at (c) $s = 0.25$,
(d) $s = 0.5$, and (e) $s = 0.75$.
 \label{fig:ChebProjectionFigure}}
           \end{center}
         \vspace{-.10in}
        \end{figure}

A general periodic curvature function can be represented as a Fourier
series in $x = s + U_w t$. For any such function, with amplitude
scaled appropriately,
\cite{AlbenSnake2013II} predicts that
the efficiency should tend to the optimum as $U_w \to 0$. For simplicity
we focus on a single mode $\sin(2\pi x) = \sin(2\pi s/U_w + 2 \pi t)$
and ask: how well can it be approximated by the truncated Chebyshev series
we have used in the numerics? At any time $t$,
$\sin(2\pi s/U_w + 2 \pi t)$ can be decomposed
into a linear combination of $\sin(2\pi s/U_w)$ and $\cos(2\pi s/U_w)$,
so we compute the best approximations to these two functions in $L^\infty$
norm by Chebyshev series with $n_1$ terms. In figure
\ref{fig:ChebProjectionFigure} we show the resulting approximation
errors (in $L^2$ and $L^\infty$) for $n_1 = 5$ and 10,
and for $\sin(2\pi s/U_w)$ (panel a) and
$\cos(2\pi s/U_w)$ (panel b). We find the error transitions
from small to large for the wavelength $U_w$ in the neighborhood of
1 for $n_1 = 5$ and 1/2 for $n_1 = 10$. For comparison, in panels c-e
we plot snapshots of the curvatures from the optimal snake
motions identified previously for $\mu_t = 30$ with
$m_1 = n_1 = 5$ and $m_1 = n_1 = 10$. The panels show the snapshots
at instants when the curvatures have local maxima at $s = 0.25$ (c),
0.5 (d) and 0.75 (e). These shapes are approximately sinusoidal, and
those with $m_1 = n_1 = 10$ have a somewhat higher wavelength
than those with $m_1 = n_1 = 5$. The wavelengths for
$m_1 = n_1 = 5$ appear to be somewhat larger than 1, while those
for $m_1 = n_1 = 10$ appear to be approximately 1.5. Based on
panels a and b, we would have expected the $m_1 = n_1 = 10$ cases in c-e
to have somewhat shorter wavelengths, but in the space spanned by
the finite number of Chebyshev polynomials, the smallest $U_w$ is not
the only consideration: certain functional forms for the curvature
should be expected to perform better than others, and the results
here may tend towards such forms.

\noindent\section{Supplementary Material:
Optimal snake motions across the $\mu_t$ transition \label{sec:modmut}}

\begin{figure} [h]
           \begin{center}
           \begin{tabular}{c}
               \includegraphics[width=5in]{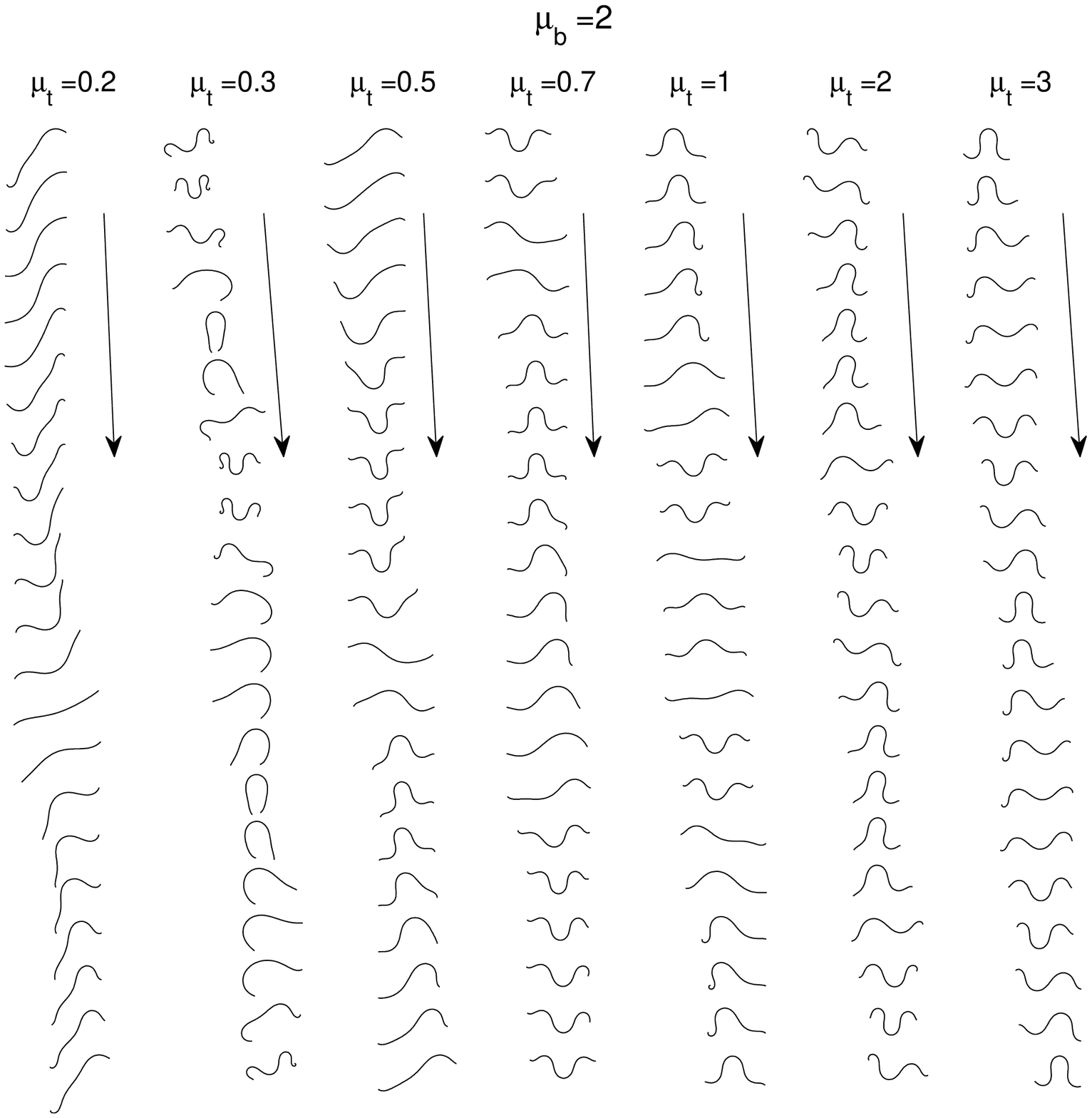} \\
           \vspace{-.15in}
\includegraphics[width=5in]{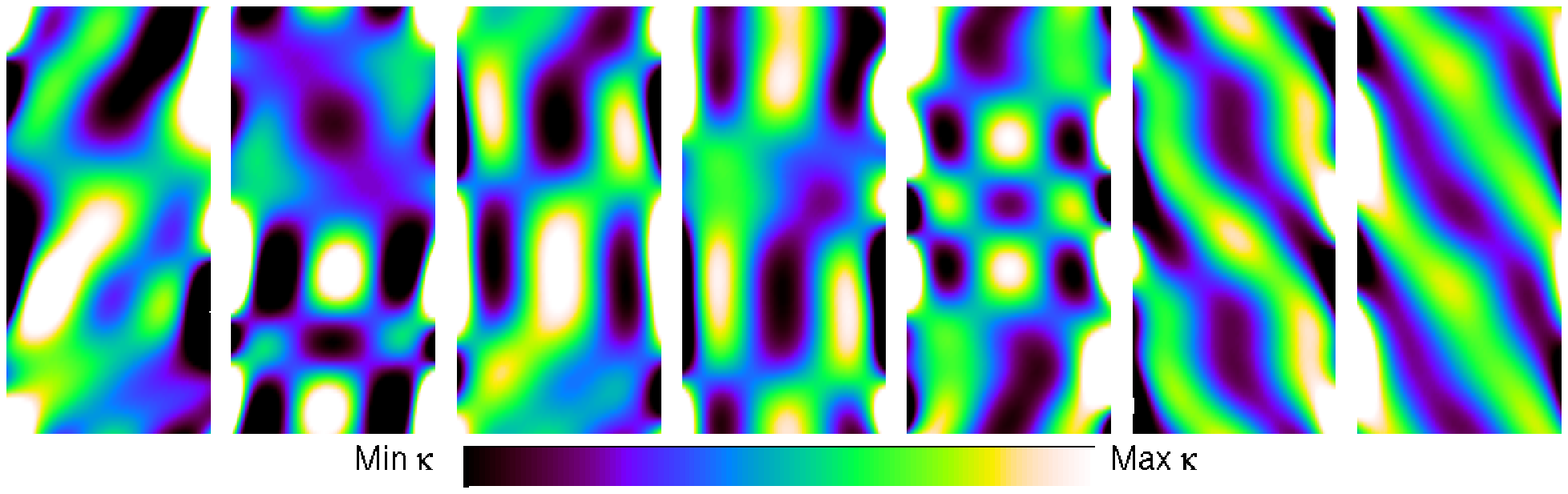} \\
           \end{tabular}
          \caption{\footnotesize Snapshots of optimal snake motions at
various $\mu_t$ (one per column, labeled at the top), for $\mu_b = 2$.
 \label{fig:Multi2}}
           \end{center}
         \vspace{-.0in}
        \end{figure}

Figure \ref{fig:Multi2} shows optima for
$\mu_b = 2$, in the same format as figure
\ref{fig:Multi1},
at a different set of seven $\mu_t$. The
motions occur on a horizontal line in figure
\ref{fig:PhasePlaneFig} which touches the bottom of
the region in which standing waves occur. Starting in the right
column, which has $\mu_t = 3$, we see a
retrograde wave, with modulations (or bulges) in the bands of
the curvature plot. In the next five columns, we see the
transition from retrograde to direct waves. However,
the curvature plots begin to resemble a checkerboard pattern.
This resemblance is incomplete, and the motions are still
retrograde or direct waves, but not standing waves.

\begin{figure} [h]
           \begin{center}
           \begin{tabular}{c}
               \includegraphics[width=5in]{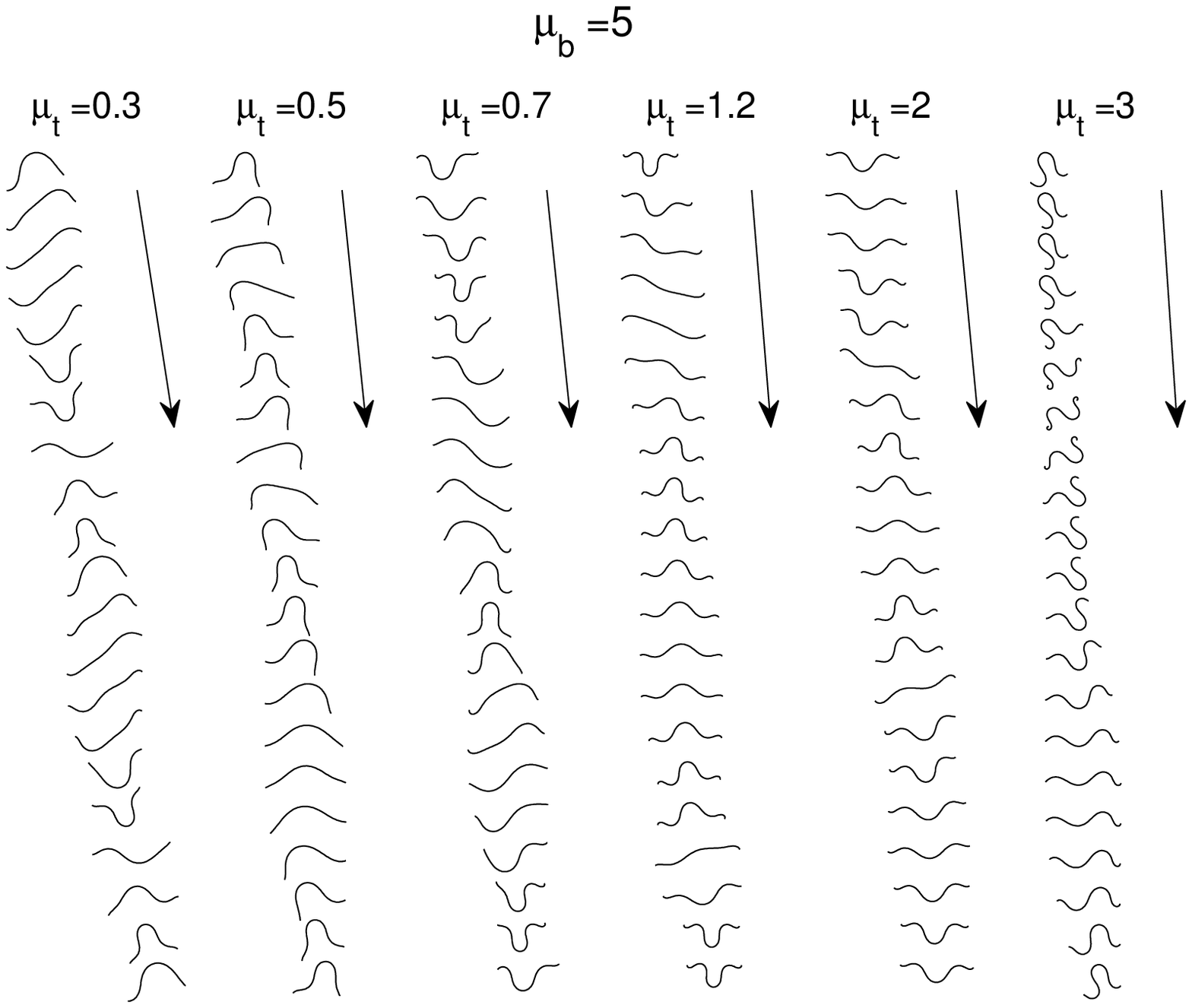} \\
           \vspace{-.15in}
\includegraphics[width=5in]{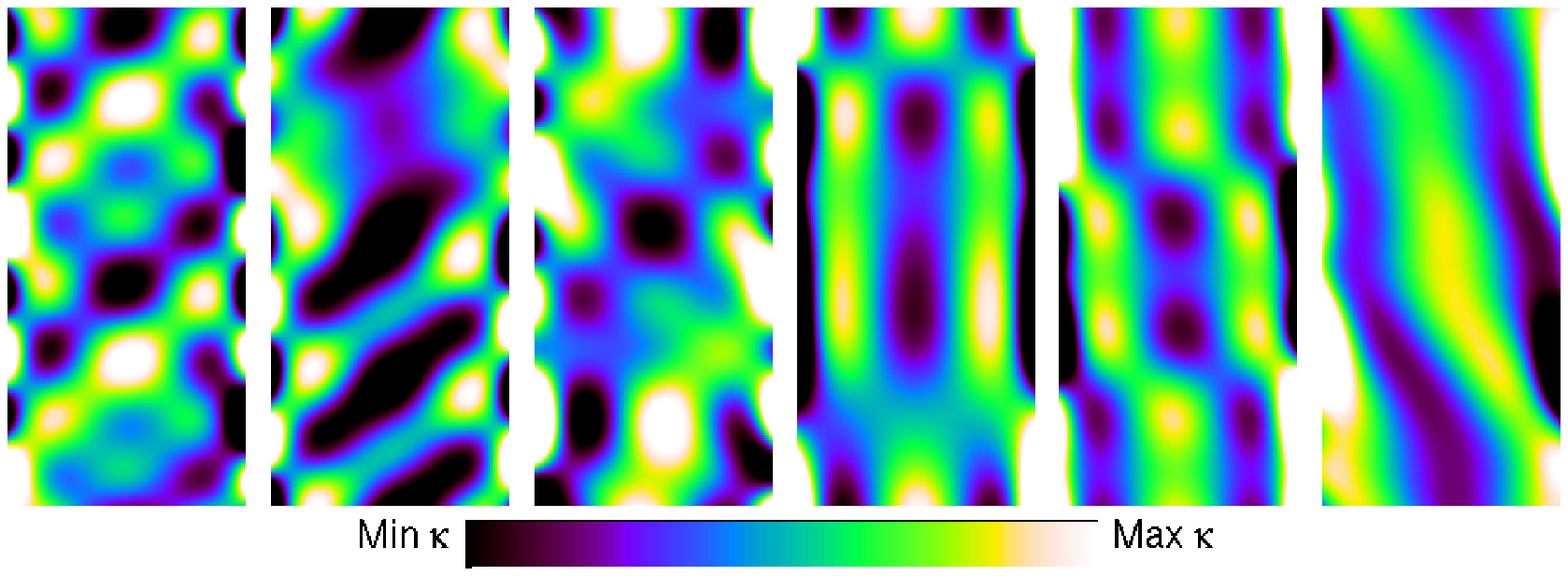} \\
           \end{tabular}
          \caption{\footnotesize Snapshots of optimal snake motions at
various $\mu_t$ (one per column, labeled at the top), for $\mu_b = 5$.
 \label{fig:Multi5}}
           \end{center}
         \vspace{-.0in}
        \end{figure}

Figure \ref{fig:Multi5} shows optima in for
$\mu_b$ increased to 5. The retrograde wave is somewhat
difficult to discern at $\mu_t = 3$, without consulting
the curvature plot. The motions are qualitatively similar
to those for $\mu_b = 3$, shown in figure \ref{fig:Multi3}.

\begin{figure} [h]
           \begin{center}
           \begin{tabular}{c}
               \includegraphics[width=5in]{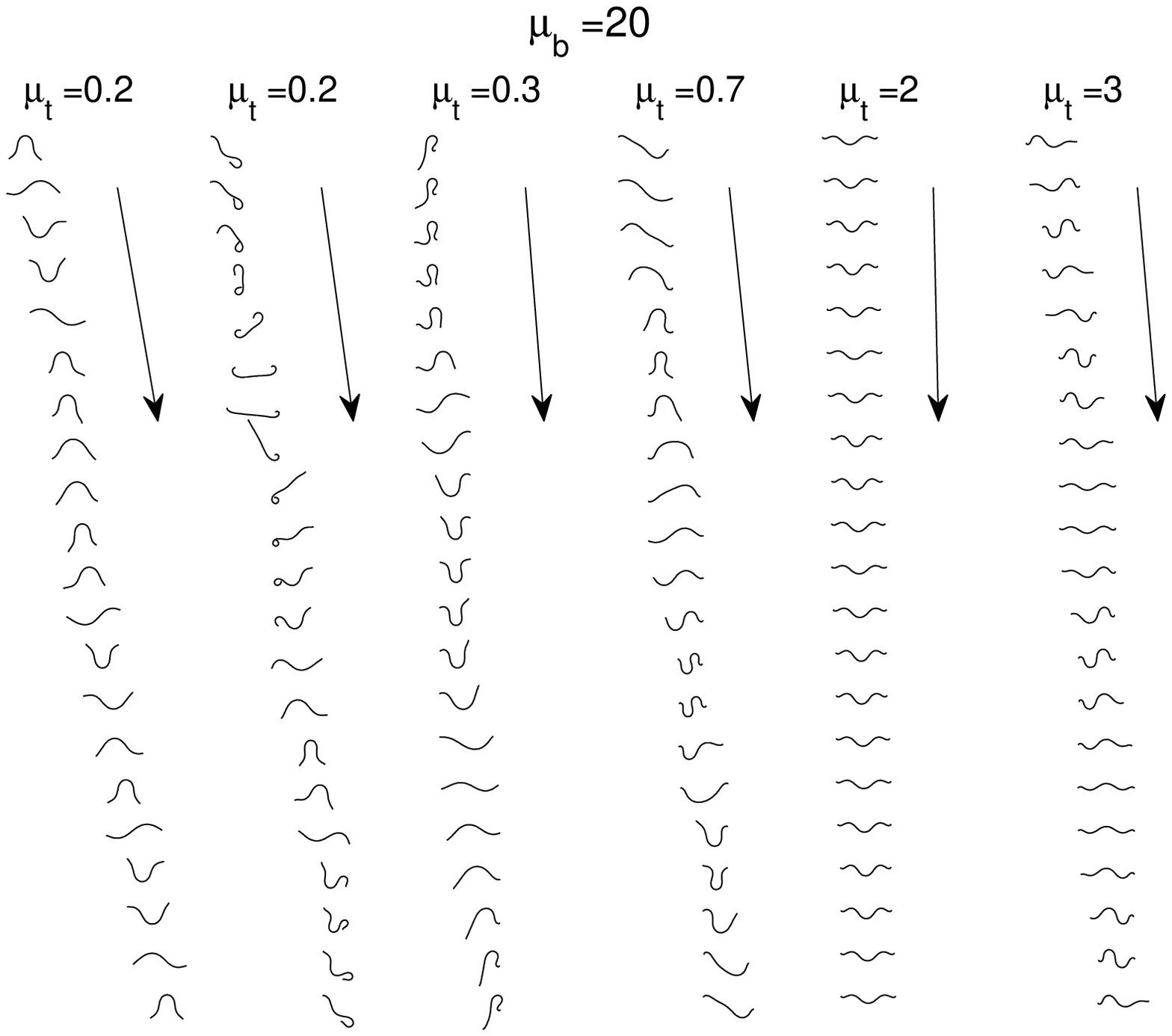} \\
           \vspace{-.15in}
\includegraphics[width=5in]{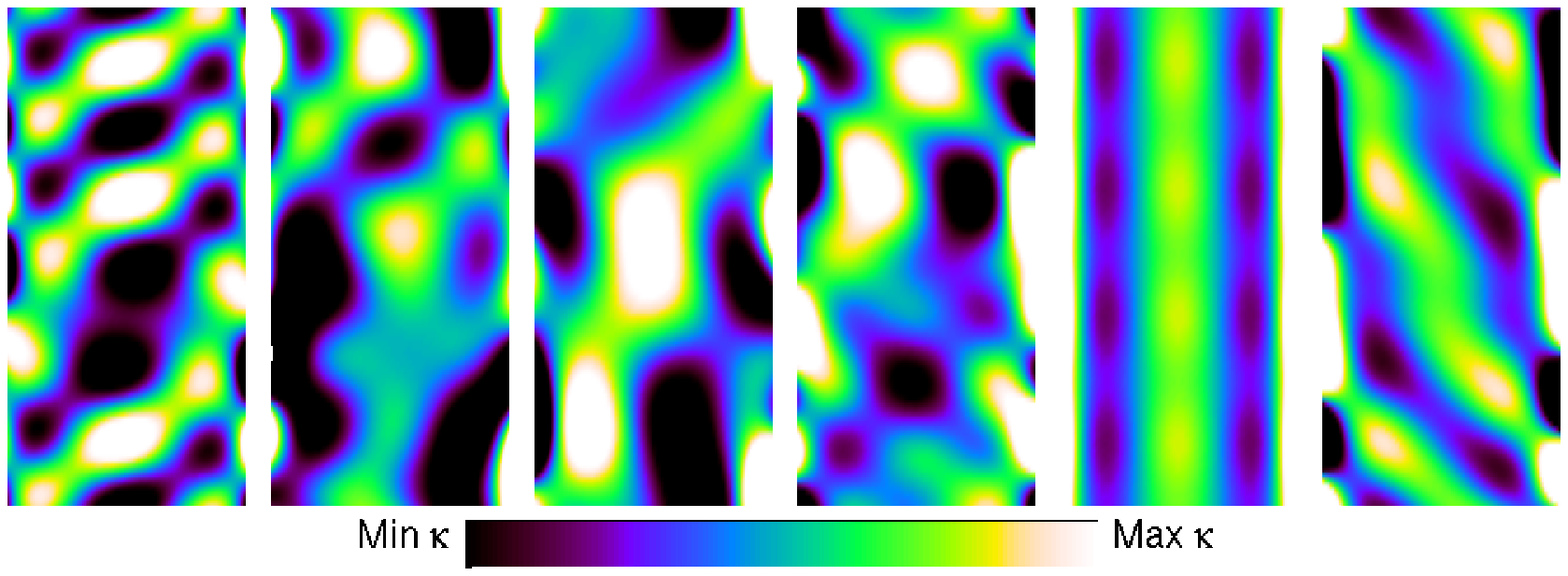} \\
           \end{tabular}
          \caption{\footnotesize Snapshots of optimal snake motions at
various $\mu_t$ (one per column, labeled at the top), for $\mu_b = 20$.
 \label{fig:Multi20}}
           \end{center}
         \vspace{-.0in}
        \end{figure}

Figure \ref{fig:Multi20} shows optima in for
$\mu_b$ increased to 20. For $\mu_t = 0.2$ we show two local optima.
The second column shows a ``swing-and-anchor'' motion. The snake
forms a loop at one end, the ``anchor,'' and then the rest of the body
swings forward, rotating about the anchor. The motion also
shows, briefly, a direct wave. This example shows that optima
can combine multiple types of motions within a single period of motion.
We have seen many other such ``mixed'' optima at moderate $\mu_t$,
but they are not reproduced here for brevity.
The other $\mu_t = 0.2$ optimum, shown in the first column,
is essentially a direct wave. The optima for $\mu_t = 0.3$,
0.7, and 3 also involve the formation of ``anchors'' near the endpoints,
somewhat similar to the concertina motion of biological snakes
\citep{MaHu2012a}. The optimum at $\mu_t = 2$ is a very symmetrical
ratcheting motion.

\noindent\section{Supplementary Material:
Triangular wave motion across $\mu_t$ \label{sec:trimut}}

\begin{figure} [h]
           \begin{center}
           \begin{tabular}{c}
               \includegraphics[width=5in]{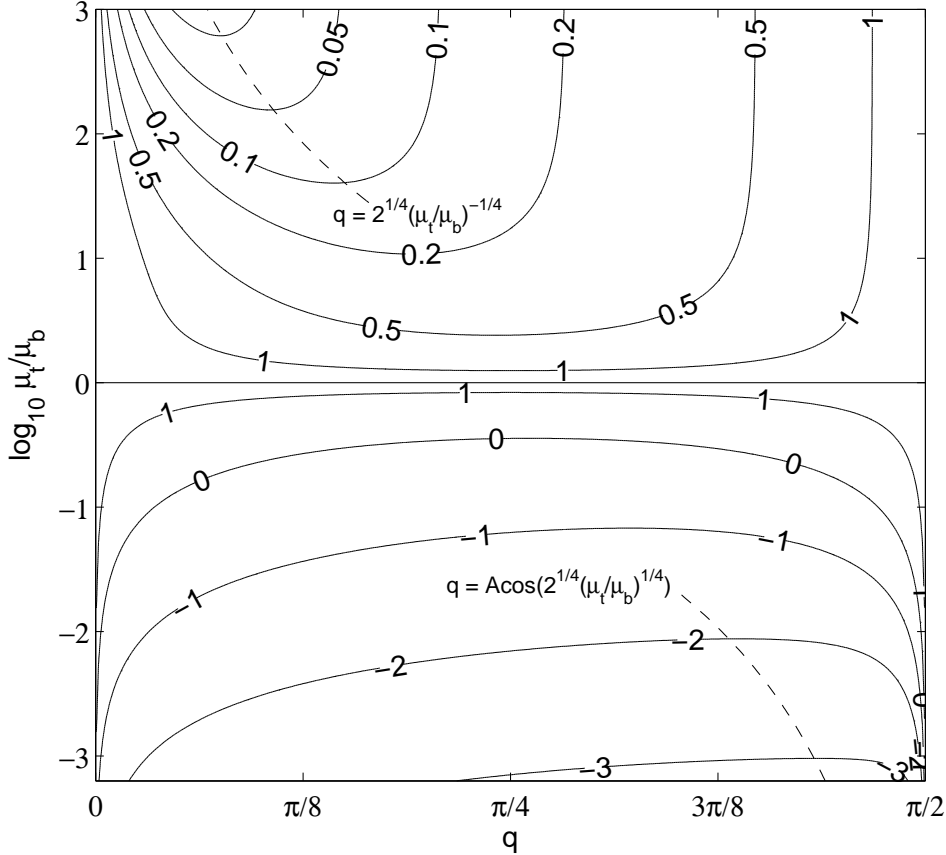} \\
           \vspace{-.25in}
           \end{tabular}
          \caption{\footnotesize Contour plot of $\log_{10} \eta/\mu_b$
as a function of the triangular wave angle $q = \arcsin{A_0}$ and
the friction coefficient ratio $\log_{10} \mu_t/\mu_b$.
 \label{fig:TriangleWaveFig}}
           \end{center}
         \vspace{-.10in}
        \end{figure}

We again consider the triangular wave
motion (\ref{y})--(\ref{dtx}), but now $\mu_t$ is nonzero,
so (\ref{fxalt}) takes the more general form
\begin{align}
\int  \mathcal{H} &\widehat{\partial_t{\mathbf{X}}}\cdot \hat{\mathbf{s}} \,s_x
+ \mu_t \widehat{\partial_t{\mathbf{X}}}\cdot \hat{\mathbf{n}} \,n_x\, ds = 0, \label{fxalt1} \\
\mathcal{H} &\equiv \left(\mu_f H(\widehat{\partial_t{\mathbf{X}}}\cdot \hat{\mathbf{s}})
+ \mu_b (1-H(\widehat{\partial_t{\mathbf{X}}}\cdot \hat{\mathbf{s}}))\right)
\end{align}
\nn with $\hat{\mathbf{s}}$ as in (\ref{stri}) and
\bqe
\hat{\mathbf{n}} = \left( {\begin{array}{c} -A_0 \mbox{sgn}(\sin(2\pi (s-t))) \\ \sqrt{1-A_0^2} \end{array} } \right). \label{ntri}
\eqe
We solve (\ref{fxalt1}) for $U$:
\bqe
U = -\frac{(-\mathcal{H} + \mu_t) A_0^2 \sqrt{1-A_0^2}}
{\mathcal{H} + (\mu_t-\mathcal{H})A_0^2}. \label{Ugen}
\eqe
\nn When $\mu_t = 0$, (\ref{Ugen}) becomes (\ref{Usol}).
The tangential speed is
\bqe
u_s = U s_x + \partial_t y s_y =
-\frac{A_0^2 \mu_t}{\mathcal{H} + (\mu_t-\mathcal{H})A_0^2} \leq 0,
\eqe
\nn so $\mathcal{H} = \mu_b$. The snake speed (\ref{Ugen}), as well
as $u_s$ and $u_n$, are
functions of $A_0$ and $\mu_t/\mu_b$ only.
By (\ref{W}) and (\ref{eta}), so is $\eta/\mu_b$.  In figure
\ref{fig:TriangleWaveFig} we give a contour
plot of $\log_{10} \eta/\mu_b$ as a function of
$q = \arcsin{A_0}$ and $\log_{10} \mu_t/\mu_b$. $q$ is the angle
that the triangle wave makes with the horizontal axis.

The forward speed $U$ is zero, and the cost of locomotion infinite,
when $\mu_t = \mu_b$, represented by a horizontal line near the middle
of figure \ref{fig:TriangleWaveFig}. This is an example
of the ``singular point''
mentioned in the previous section, when $\mu_t = \mu_b = \mu_f = 1$.
When $\mu_t > \mu_b$, above
the horizontal line, $U < 0$, and the snake moves as a retrograde wave. $\eta/\mu_b$
decays to 1 as $\mu_t/\mu_b \to \infty$. In this regime, for fixed
$\mu_t$,
$\eta$ itself is the product of $\mu_b$ with a
function that decays more slowly than $\mu_b$ as $\mu_b \to 0$. Thus $\eta$ is
minimized when $\mu_b \to 1$, the minimal value of $\mu_b$. We plot the
optimal amplitude in the limit of large $\mu_t/\mu_b$
(derived for a wave of any shape in \cite{AlbenSnake2013II}) as a dashed line at the upper left of the figure,
with the formula for the optimal amplitude in this limit.
In the lower portion of figure \ref{fig:TriangleWaveFig},
$\mu_t < \mu_b$, $U > 0$, and the snake moves as a direct wave.
Here $\eta/\mu_b$ decays to 0 as $\mu_t/\mu_b \to 0$.
In this regime, for fixed $\mu_t$,
$\eta$ itself is the product of $\mu_b$ with a
function that decays more rapidly than $\mu_b^{-1}$ as $\mu_b \to \infty$.
Thus $\eta$ is minimized by taking $\mu_b \to \infty$ and
$\mu_t \to 0$. The wave angle which minimizes $\eta$ can be found
by directly minimizing $\eta$ with respect to $q$. We find that in
the limit $\mu_t/\mu_b \to 0$, the optimal wave angle
tends to $\pi/2$, i.e. the optimal wave becomes steeper.

\end{document}